\documentclass{article}

\usepackage{amsfonts,amsmath,amssymb}
\usepackage{hyperref}
\usepackage{color}
\usepackage{graphicx}
\usepackage{subcaption}
\usepackage[section]{placeins}
\usepackage[utf8]{inputenc}

\definecolor{darkblue}{rgb}{0,0.08,0.45}
\hypersetup{colorlinks,linkcolor=darkblue,citecolor=darkblue,filecolor=darkblue,urlcolor=darkblue}
\hypersetup{bookmarksopen,bookmarksnumbered}

\renewcommand{\d}{\ensuremath{\,\mathrm{d}}}

\title{Modeling frequency shifts of collective bubble resonances with the boundary element method\footnote{Copyright (2023) Acoustical Society of America. This article may be downloaded for personal use only. Any other use requires prior permission of the author and the Acoustical Society of America. The following article appeared in The Journal of the Acoustical Society of America, volume 153, issue 3, pages 1898--1911, March 2023, and may be found at \url{https://doi.org/10.1121/10.0017650}.}}

\author{Rudyard Jerez Boudesseul\thanks{Institute for Mathematical and Computational Engineering, School of Engineering and Faculty of Mathematics, Pontificia Universidad Católica de Chile, Santiago, Chile.\newline Contact: e.wout@uc.cl} \and Elwin van 't Wout\footnotemark[2]}

\date{March 22, 2023}

\begin{document}
	
\maketitle

\begin{abstract}
Increasing the number of closely-packed air bubbles immersed in water changes the frequency of the Minnaert resonance. The collective interactions between bubbles in a small ensemble are primarily in the same phase, causing them to radiate a spherically-symmetric field that peaks at a frequency lower than the Minnaert resonance for a single bubble. In contrast, large periodic arrays include bubbles that are further apart than half the wavelength, so that collective resonances have bubbles oscillating in opposite phases, ultimately creating a fundamental resonance at a frequency higher than the single-bubble Minnaert resonance. This work investigates the transition in resonance behavior using a modal analysis of a mass-spring system and a boundary element method. We significantly reduce the computational complexity of the full-wave solver to a linear dependence on the number of bubbles in a rectangular array. The simulated acoustic fields confirm the initial downshift in resonance frequency and the strong influence of collective resonances when the array has hundreds of bubbles covering more than half the wavelength. These results are essential in understanding the low-frequency resonance characteristics of bubble ensembles, which have important applications in diverse fields such as underwater acoustics, quantum physics, and metamaterial design.
\end{abstract}

\section{Introduction}
\label{sec:introduction}

Gas-filled bubbles immersed in a fluid exhibit interesting dynamics when excited by acoustic pressure fields. For example, the resonance behavior of bubbles at low frequency causes a strong, spherically-symmetric scattered field, which has been known since the early work of Minnaert~\cite{minnaert1933on} and well studied with analytical expressions~\cite{anderson1950sound}. The acoustic interactions between multiple bubbles create different resonances, with a monopole resonance at a lower frequency and sharp superresonances at higher frequencies than the Minnaert resonance~\cite{feuillade1995scattering}. These resonance phenomena of bubble ensembles are known as collective resonances~\cite{zeravcic2011collective}, or proximity resonances~\cite{heller1996quantum}, and play an important role in diverse fields such as underwater acoustics~\cite{raveau2016resonance}, metamaterial design~\cite{lanoy2017acoustic}, and quantum physics~\cite{gaspard2022resonance}. They have been detected in field~\cite{kolaini1994low, diachok2001interpretation, leifer2007acoustic} and laboratory~\cite{loewen1991model, rathgen2007nanometer, aldham2010measurement, greene2012laboratory, harazi2019acoustics, combriat2020acoustic, boughzala2021polyhedral} experiments, and studied using various mathematical approaches, such as homogenization into effective media~\cite{hahn2007low}, modal analysis of ordinary differential equations for mass-spring systems~\cite{manasseh2009frequencies}, linearized Rayleigh-Plesset equations~\cite{devaud2008minnaert}, and numerical methods for the Helmholtz equation~\cite{wout2021proximity}.

The Minnaert resonance of a single air-filled bubble immersed in water at atmospheric pressure has a frequency of $ka \approx 0.014$, where $k$ is the wavenumber, and $a$ is the bubble radius, indicating that the bubble radius is hundreds of times smaller than the wavelength. The interaction between two closely separated bubbles, or placing a bubble close to a wall, can cause either a symmetric resonance, where both bubbles oscillate in phase, or an antisymmetric superresonance, with oscillations in opposite phases, depending on the boundary conditions~\cite{manasseh2009frequencies}. The primary resonance in the symmetric case is at a lower frequency, whereas the superresonance has a sharp peak at a frequency up to two times the resonance frequency of a single uncoupled bubble~\cite{wout2021proximity}. These characteristics extend to small bubble ensembles, where the dominant resonance displays a downward frequency shift, and where several sharp superresonances occur at higher frequencies~\cite{feuillade1995scattering}. Experiments using millimeter-sized air-filled bubbles in water tanks confirm this behavior for cloud~\cite{leroy2005bubble} and chain~\cite{nikolovska2007propagation, roshid2020extraction} configurations, bubbles close to a wall~\cite{payne2005symmetric, illesinghe2009eigenmodal}, and bubbles trapped in a yield-stress fluid~\cite{lanoy2015manipulating}.

In contrast to small ensembles, screens of bubbles organized in a periodic array show a dominant resonance at a frequency higher than the Minnaert resonance, as observed in laboratory experiments with metascreens~\cite{leroy2015superabsorption, lanoy2018broadband} and bubble arrays in a yield-stress fluid~\cite{leroy2009transmission, leroy2018acoustics}. These experimental results match well with mathematical models based on monopole resonances of mass-spring systems~\cite{leroy2009transmission}, effective media~\cite{sharma2020sound}, and finite element methods~\cite{calvo2015underwater, leroy2018acoustics} for infinite periodic arrays. The dynamics of these large bubble arrays exhibit exciting features for metamaterial design such as superabsorption~\cite{leroy2015superabsorption}, metascreen coatings~\cite{shankar2022acoustic}, subwavelength focusing~\cite{lanoy2015subwavelength}, and phononic band gaps~\cite{leroy2009design}.

This manuscript studies the transition zone between small ensembles and periodic arrays, where the fundamental resonance frequency changes from a downshift to an upshift compared to a single bubble. While the collective resonances of either a few bubbles or infinite arrays are well documented, the resonance frequencies at intermediate ensemble sizes have attracted little attention in the literature. The phenomenon of a changing frequency shift at large, but finite, arrays has been attributed to scattering interactions between closely spaced bubbles~\cite{calvo2015underwater}, more specifically, bubbles oscillating in opposite phases in rings of half a wavelength~\cite{leroy2009transmission}, with an aggregate effect that increases the resonance frequency~\cite{leroy2018acoustics}. Here, we investigate the frequency shifts of collective bubble resonances using mathematical models. The modal analysis of a mass-spring model indicates that the fundamental resonance frequency decreases, then reverses, and finally converges to a frequency higher than Minnaert's when enough bubbles are added to the ensemble. Moreover, higher-order collective resonances are excited significantly by an incident plane wave traveling in the direction normal to the array. Solving the Helmholtz equation using the boundary element method (BEM) supports these observations. In addition, this full-wave solver shows the acoustic field of a square array with hundreds of bubbles oscillating in phase at the fundamental resonance, and in opposite phases at the collective resonances.

Simulating large bubble arrays using the full-wave solver is feasible after developing a custom acceleration technique for rectangular arrays of bubbles, reducing the BEM's computational complexity to a linear dependence on the bubble count, regardless of their separation distances. Unlike volumetric full-wave solvers, such as the finite element method, the BEM solves a boundary integral equation and does not mesh the space between the bubbles. It uses a triangular grid at the bubble surfaces and naturally handles the acoustic radiation for an unbounded exterior domain. Furthermore, we exploit the translation invariance of Green's functions to compress the all-to-all interaction between the bubbles into an efficient matrix-free implementation of the BEM. These accelerations allow fast frequency sweeps and accurate simulations for square arrays containing hundreds of bubbles.

The following section presents the equations of motion for the bubble dynamics, including the mass-spring model and the Helmholtz equation. The modal analysis of the mass-spring system of large bubble ensembles will be presented in Section~\ref{sec:formulation:mass_spring}, and the accelerated BEM for acoustic transmission in Section~\ref{sec:formulation:bem}. Section~\ref{sec:results} presents the numerical results of the resonance modes of the mass-spring system, together with the scattered field of large bubble arrays calculated using the BEM.

\section{Formulation}
\label{sec:formulation}

The dynamics of an air bubble immersed in a fluid depend strongly on the frequency of the acoustic wave field. This study considers low-frequency wave fields, where $0 < a \ll \lambda$, where $a$ is again the bubble radius, and $\lambda$ is the wavelength of the incident field. In this case, the bubble exhibits a breathing motion, expanding and contracting uniformly. We use two mathematical models: a mass-spring system of ordinary differential equations, and the BEM for the Helmholtz equation.
The incident wave field is a plane wave
\begin{equation}
	\label{eq:pinc}
	p_{\mathrm{inc}}(\mathbf{x}, t) = P_\mathrm{inc} e^{\imath(k \mathbf{x} \cdot \mathbf{v}_\mathrm{inc} - \omega t)}
\end{equation}
at location $\mathbf{x} \in \mathbb{R}^3$ and time $t$, where $P_\mathrm{inc}$ denotes the amplitude, $\mathbf{v}_\mathrm{inc}$ the propagation direction, $k$ the wavenumber, $\omega$ the angular frequency, and $\imath$ the imaginary unit. All propagating media are modeled as acoustic fluids for which shear waves~\cite{skvortsov2019sound, sharma2020superscattering} can be neglected.

\subsection{Mass-spring model}
\label{sec:formulation:mass_spring}

The mass-spring model assumes the bubble to be a monopole point scatterer~\cite{feuillade1995scattering}, which is accurate for the low-frequency fields considered in this study and allows for analytical expressions of resonance characteristics.

\subsubsection{Single bubble}

The equation of motion for a single bubble located at the global origin is given by
\begin{equation}
	\label{massSpringEquation}
	m \nu''(t) + b \nu'(t) + \kappa\nu(t) = -P_\mathrm{inc} e^{-\imath \omega t},
\end{equation}
which is an ordinary differential equation representing a mass-spring system for $\nu$ the unknown bubble volume differential~\cite{leighton2012acoustic}. Notice that this variable depends only on time, since the bubble is assumed to be a point scatterer with monopole dynamics~\cite{medwin1997bubbles}. Furthermore, $m = \rho/(4\pi a)$ denotes the equivalent mass, $\rho$ the fluid density, $\kappa = 4.2 P_A / (4 \pi a^3)$ the bubble stiffness, $P_A$ the ambient pressure, $b = \kappa a/c$ the damping, and $c$ the speed of sound in the fluid. Here, the parameter~$b$ includes radiation damping~\cite{spratt2017radiation} only, which is relatively small compared to the bubble motion at resonance: $0 < b \ll m\omega_0$.

Considering harmonic wave field solutions
\begin{equation}
	\nu(t) = \bar{\nu} e^{-\imath \omega t}
\end{equation}
leads to a solution of the mass-spring equation~\eqref{massSpringEquation} for $\bar{\nu}$, i.e.,
\begin{equation}
	\label{nu_expression}
	\bar{\nu} = \frac{ -\frac{P_\mathrm{inc}}{m\omega^2} }{ \frac{\omega_0^2}{\omega} - 1 - \imath \delta },
\end{equation}
where $\delta = b/(m\omega)$ denotes a non-dimensional damping parameter, and $\omega_0 = \sqrt{\kappa/m}$ is the natural resonance frequency~\cite{leighton2012acoustic}. Since damping is relatively small ($0 < \delta \ll 1$), a sharp peak in bubble amplitude is obtained near $\omega_0$, which is also known as the Minnaert resonance frequency~\cite{ainslie2011review}. The quality factor $Q = \omega_0/(\omega_+ - \omega_-)$ characterizes the width of the resonance peak, where $\omega_\pm$ are the frequencies at which the pressure amplitude corresponds to the half-power points compared to the resonance peak~\cite{devin1959survey}. The quality factor depends on the damping parameter as
\begin{equation}
	Q = \frac{\sqrt{1 - \delta^2}}{\sqrt{1 + \delta} - \sqrt{1 - \delta}}.
\end{equation}
Notice that $Q \approx 1/\delta$, since $\delta \ll 1$.

\subsubsection{Multiple bubbles}

Consider a configuration with $N$ bubbles of identical mass, damping, stiffness, and radius. Their centers are located at $\mathbf{c}_i \in \mathbb{R}^3$, with separation distances $d_{ij} = |\mathbf{c}_i - \mathbf{c}_j| > 2a$, for $i,j = 1,2,3,\dots,N$. The same mass-spring system~\eqref{massSpringEquation} models the dynamics of each bubble~\cite{feuillade1995scattering}, that is,
\begin{equation}
	\label{NBubblesAnalysis}
	m \nu''_i(t) + b \nu'_i(t) + \kappa \nu_i(t) = -f_i(t) - \sum_{j=1}^N \rho g_{ij} \nu''_j(t), \quad \text{for } i = 1,2,3,\dots,N;
\end{equation}
where
\begin{align}
	f_i &= P_\mathrm{inc} e^{\imath (k \mathbf{c}_i \cdot \mathbf{v}_\mathrm{inc} - \omega t)}; \\
	g_{ij} &= \begin{cases} e^{\imath k d_{ij}}/(4 \pi d_{ij})& \text{for } i \ne j, \\ 0 & \text{for } i = j. \end{cases} \label{eq:green:mass:spring}
\end{align}
The extra sum on the right-hand side comes from the self-consistent approach of adding the field scattered from all other bubbles to the incident wave field~\cite{manasseh2009frequencies}. This linear system of coupled ordinary differential equations allows a modal analysis of the bubble dynamics. Here, we define a \emph{mode} as a linear combination of bubble movements
\begin{equation}
	\label{eq:mode:def}
	\mu(t) = \sum_{i=1}^N a_i \nu_i(t)
\end{equation}
for coefficients $a_i \in \mathbb{C}$, that satisfies a decoupled differential equation
\begin{equation}
	\label{eq:massspring:mode}
	m \mu''(t) + b \mu'(t) + \kappa \mu(t) = -\sum_{i=1}^N a_i f_i(t) -
	\rho \sigma \mu''(t)
\end{equation}
for $\sigma \in \mathbb{C}$. Because of the linearity of the differential operators, the modes need to satisfy the condition
\begin{equation}
	\sum_{j=1}^N \sum_{i=1}^N a_i g_{ij} \ddot{\nu}_j(t) = \sum_{j=1}^N \sigma a_j \ddot{\nu}_j(t),
\end{equation}
which is equivalent to the eigenvalue problem
\begin{equation}
	\label{eq:greenmatrix:spectrum}
	G \mathbf{a} = \sigma \mathbf{a}
\end{equation}
whose elements $(G)_{ij} = g_{ij}$ are Green's functions~\eqref{eq:green:mass:spring}. Green's matrix~$G$ is complex-valued and symmetric. It is non-normal except for rare bubble configurations. Hence, eigenvalues can be complex-valued and degenerate, and eigenvectors are not necessarily orthogonal. Given an eigenpair $(\sigma_l, \mathbf{a}_l)$, the harmonic function
\begin{equation}
	\mu_l(t) = \bar\mu_l e^{-\imath \omega t}
\end{equation}
solves the modal differential equation~\eqref{eq:massspring:mode} when
\begin{equation}
	\label{eq:barmu:solution}
	\bar\mu_l = \frac{ -\frac{\tilde{f}_l}{\tilde{m}_l\omega^2} }{ \frac{\tilde\omega_l^2}{\omega} - 1 - \imath \tilde\delta_l }
\end{equation}
for $l=1,2,\dots,N$.
Here,
\begin{align}
	\tilde{f}_l &= P_\mathrm{inc} \sum_{j=1}^N a_{l,j} e^{\imath k \mathbf{c}_j \cdot \mathbf{v}_\mathrm{inc}}, \\
	\tilde{m}_l &= m + \rho \Re(\sigma_l), \\
	\tilde{b}_l &= b + \omega \rho \Im(\sigma_l), \\
	\tilde{\delta}_l &= \tilde{b}_l/(\omega \tilde{m}_l), \\
	\tilde{\omega}_l &= \sqrt{\kappa/\tilde{m}_l}, \label{eq:tilde:omega} \\
	\tilde{Q}_l &= \sqrt{1 - \tilde\delta_l^2} \Big/ \left(\sqrt{1 + \tilde\delta_l} - \sqrt{1 - \tilde\delta_l}\right) \label{eq:tilde:q}
\end{align}
are the parameters that describe each mode. The eigenvalues depend on the homogeneous differential equation at frequency~$\omega$, but the incident wave field forces each mode differently~\cite{zeravcic2011collective}. For example, superresonances with pairs of bubbles oscillating in opposite phases will not be excited by a perpendicular plane wave. To quantify the relative forcing of each mode, let us define the \emph{excitation factor} of mode~$l$ as
\begin{equation}
	\label{eq:excitationfactor}
	\hat{f}_l = \frac{1}{P_\mathrm{inc} \sqrt{N}} |\tilde{f}_l| = \frac1{\sqrt{N}} \left| \sum_{j=1}^N a_{l,j} \right|
\end{equation}
where $\mathbf{c}_j \cdot \mathbf{v}_\mathrm{inc} = 0$ for bubbles located in an array perpendicular to the propagation direction of the incident plane wave~\eqref{eq:pinc}. Since all eigenvectors~$\mathbf{a}_l$ are normalized, the value of $\hat{f}_l$ lies between zero and one.

\subsubsection{Infinite bubble array}

A full modal analysis becomes impracticable for periodic arrays of bubbles since an infinite number of modes are present. In the special case where all bubbles are in a plane perpendicular to the excitation direction, one can assume that all bubbles oscillate with identical amplitude and phase at the fundamental mode~\cite{weston1966acoustic}. This is equivalent to considering a constant eigenvector for the infinite Green's matrix~\eqref{eq:greenmatrix:spectrum}. By construction,
\begin{equation}
	\sigma_\infty = \sum_j G_{ij} = \sum_{j \ne i} \frac{e^{\imath k d_{ij}}}{4 \pi d_{ij}}
\end{equation}
for an arbitrary `center' bubble~$i$. This result is consistent with mass-spring theory for periodic arrays~\cite{leroy2009transmission}, and limits of the infinite series are known for cases such as chains and square arrays~\cite{tolstoy1990line}.

\subsection{Acoustic wave propagation model}
\label{sec:formulation:bem}

The Helmholtz equation accurately models harmonic acoustic wave propagation and, unlike the mass-spring approach, includes the actual three-dimensional shape of the bubbles. On the downside, solving the system of partial differential equations using numerical methods, such as finite differences, finite elements, or boundary elements, is computationally expensive for large-scale configurations~\cite{steinbach2008numerical}.

\subsubsection{Helmholtz equation}

Let us denote the volume of each bubble by $\Omega_j \subset \mathbb{R}^3$ and its surface by $\Gamma_j$ for $ j = 1,2,3,\dots,N$. The boundaries $\Gamma_j$ have a normal vector $\hat{\mathbf{n}}_j$ of unit length and point outward from bubble~$j$. The bubbles are disjoint, so that $\Omega_i \cap \Omega_j = \emptyset$, $\forall i \ne j$.
Furthermore, we define the variables $\Omega = \bigcup\limits_{j=1}^{N} \Omega_j$ and
$\Gamma = \bigcup\limits_{j=1}^{N} \Gamma_j$ as the union of the volumes and surfaces,
and $\Omega_0 = \mathbb{R}^3\setminus\Omega$ as the unbounded exterior region. The Helmholtz system reads
\begin{equation}
	\begin{cases}
		- \Delta p_{\text{tot}} - k_q^2 p_{\text{tot}} = 0, \quad &\text{in } \Omega_q, \text{ for } q=0,1,2,\dots,N;\\
		\gamma^{+}_{\mathrm{D}, j} p_{\text{tot}} = \gamma^{-}_{\mathrm{D}, j} p_{\text{tot}}, &\text{at } \Gamma_j, \text{ for } j=1,2,\dots,N; \\
		\frac{1}{\rho_0} \gamma^{+}_{\mathrm{N}, j} p_{\text{tot}} = \frac{1}{\rho_j} \gamma^{-}_{\mathrm{N},j} p_{\text{tot}}, &\text{at } \Gamma_j, \text{ for } j=1,2,\dots,N; \\
		\lim\limits_{\mathbf{r} \to \infty} |\mathbf{r}| (\partial_{|\mathbf{r}|} p_{\text{sca}} - \imath k_0 p_{\text{sca}}) = 0;
	\end{cases}
	\label{eq:helmholtz}
\end{equation}
where $p_{\text{tot}}$ denotes the acoustic pressure, $p_{\text{sca}}$ the scattered field, and $k_i$ and $\rho_i$ the wavenumber and the density in region~$i$, respectively. The second and third equations are interface conditions for the continuity of the pressure field and normal particle velocity, respectively. The fourth equation is Sommerfeld's condition, modeling the radiation damping of the bubble dynamics. Furthermore,
\begin{equation}
	\begin{cases}
		\gamma^{-}_{\mathrm{D}, j} f(\mathbf{x}) = \lim\limits_{\Omega_j \ni \mathbf{y} \to \mathbf{x}} f(\mathbf{y}),\\
		\gamma^{+}_{\mathrm{D}, j} f(\mathbf{x}) = \lim\limits_{\Omega_0 \ni \mathbf{y} \to \mathbf{x}} f(\mathbf{y}),\\
		\gamma^{-}_{\mathrm{N},j} f(\mathbf{x}) = \lim\limits_{\Omega_j \ni \mathbf{y} \to \mathbf{x}} \nabla f(\mathbf{y}) \cdot \hat{\mathbf{n}}_j(\mathbf{x}),\\
		\gamma^{+}_{\mathrm{N}, j} f(\mathbf{x}) = \lim\limits_{\Omega_0 \ni \mathbf{y} \to \mathbf{x}} \nabla f(\mathbf{y}) \cdot \hat{\mathbf{n}}_j(\mathbf{x}),
	\end{cases}
	\label{TracesDefinition}
\end{equation}
with $\mathbf{x} \in \Gamma_j$. The operators $\gamma^+_{*,j}$ and $\gamma^-_{*,j}$ refer to the traces from the exterior or interior of bubble $j$, respectively, and $\gamma^{\pm}_{\mathrm{D},j}$ and $\gamma^{\pm}_{\mathrm{N},j}$ are the Dirichlet and Neumann traces, respectively.

\subsubsection{Acoustic boundary integral equations}

Since the bubbles and the exterior domain each have constant density and speed of sound, the scattered field can be represented as~\cite{steinbach2008numerical}
\begin{equation}
	p_{\text{sca}} = \sum\limits_{j=1}^N \left( \mathcal{V}_{0,j} \psi_j - \mathcal{K}_{0,j} \phi_j \right) \quad \text{in } \Omega_0
    \label{eq:bie:representation}
\end{equation}
where $\phi_j = \gamma_{\mathrm{D}, j}^+ p_{\text{tot}} = \gamma_{\mathrm{D}, j}^- p_{\text{tot}}$ and $\psi_j = \gamma_{\mathrm{N}, j}^+ p_{\text{tot}} = \frac{\rho_0}{\rho_j} \gamma_{\mathrm{N}, j}^- p_{\text{tot}}$ are the unknown Dirichlet and Neumann potentials at the surface of bubble~$j$, respectively. Here,
\begin{align}
	\left [\mathcal{V}_{q,j} \psi \right ](\mathbf{x}) &= \iint\limits_{\Gamma_j} G_q(\mathbf{x}, \mathbf{y}) \psi (\mathbf{y}) \d\mathbf{y} & \text{for } \mathbf{x} \in \Omega_q, \\
	[\mathcal{K}_{q,j} \phi](\mathbf{x}) &= \iint\limits_{\Gamma_j} \frac{\partial G_q(\mathbf{x}, \mathbf{y})}{\partial \hat{\mathbf{n}}_j (\mathbf{y})} \phi (\mathbf{y}) \d\mathbf{y} & \text{for } \mathbf{x} \in \Omega_q,
\end{align}
are the single-layer and double-layer potential integral operators, with
\begin{equation}
	\label{GreenFunction}
	G_q(\mathbf{x}, \mathbf{y}) = \frac{e^{\imath k_q |\mathbf{x}-\mathbf{y}|}}{4 \pi |\mathbf{x}-\mathbf{y}|} \quad \text{for } \mathbf{x}, \mathbf{y} \in \Omega_q \text{ and } \mathbf{x} \ne \mathbf{y}
\end{equation}
the Green's function of the Helmholtz equation. Different boundary integral equations formulate the Helmholtz equation as a surface potential problem at the bubble surfaces~\cite{wout2021benchmarking}. The PMCHWT formulation~\cite{poggio1973integral}
\begin{equation}
	\hat{A}_{i}
	\begin{bmatrix}
		\phi_i \\
		\psi_i \\
	\end{bmatrix}
	+
	\sum\limits_{j=1}^N
	A_{0,ij}
	\begin{bmatrix}
		\phi_j \\
		\psi_j \\
	\end{bmatrix}
	=
	\begin{bmatrix}
		\gamma_{\mathrm{D},i}^+ p_{\text{inc}}\\
		\gamma_{\mathrm{N},i}^+ p_{\text{inc}}\\
	\end{bmatrix} \quad \text{for } i=1,2,\dots,N,
	\label{PMCHWTFormulation}
\end{equation}
is one of the most robust and accurate choices for acoustics~\cite{wout2022pmchwt}. Here,
\begin{equation}
	A_{0,ij} =     
	\begin{bmatrix} 
		-K_{0,ij} & V_{0,ij}  \\
		D_{0,ij} & T_{0,ij} \\
	\end{bmatrix}
	\text{ and }
	\hat{A}_j =
	\begin{bmatrix}
		-K_{j,jj} & \frac{\rho_j}{\rho_0} V_{j,jj} \\
		\frac{\rho_0}{\rho_j} D_{j,jj} & T_{j,jj} \\
	\end{bmatrix}
	\label{calderon}
\end{equation}
denote the exterior and interior Calderón operators, respectively, where
\begin{align}
	[V_{q,ij} \psi](\mathbf{x}) &= \iint_{\Gamma_j} G_q(\mathbf{x}, \mathbf{y}) \psi (\mathbf{y}) \d\mathbf{y} & \text{for } \mathbf{x} \in \Gamma_i, \label{eq:sl} \\
	[K_{q,ij} \phi](\mathbf{x}) &= \iint_{\Gamma_j} \frac{\partial}{\partial \hat{\mathbf{n}}_j(\mathbf{y})} G_q(\mathbf{x}, \mathbf{y}) \phi(\mathbf{y}) \d\mathbf{y} & \text{for } \mathbf{x} \in \Gamma_i, \label{eq:dl} \\
	[T_{q,ij} \psi](\mathbf{x}) &= \frac{\partial}{\partial \hat{\mathbf{n}}_i (\mathbf{x})} \iint_{\Gamma_j} G_q(\mathbf{x}, \mathbf{y}) \psi(\mathbf{y}) \d\mathbf{y} & \text{for } \mathbf{x} \in \Gamma_i, \label{eq:ad} \\
	[D_{q,ij} \phi](\mathbf{x}) &= -\frac{\partial}{\partial \hat{\mathbf{n}}_i(\mathbf{x})} \iint_{\Gamma_j} \frac{\partial}{\partial \hat{\mathbf{n}}_j(\mathbf{y})} G_q(\mathbf{x}, \mathbf{y}) \phi(\mathbf{y}) \d\mathbf{y} & \text{for } \mathbf{x} \in \Gamma_i, \label{eq:hs}
\end{align}
are the single-layer, double-layer, adjoint double-layer, and hypersingular boundary integral operators, respectively. 

\subsubsection{Boundary element method}

We discretize the boundary integral equations with a Galerkin method~\cite{smigaj2015solving}. The test and basis functions are piecewise linear (P1) functions on identical triangular surface meshes for each bubble. The discretized PMCHWT formulation~\eqref{PMCHWTFormulation} reads
\begin{equation}
	\begin{bmatrix}
		\hat{\mathbf{A}}_1 + \mathbf{A}_{0,11} & \mathbf{A}_{0,12} & \cdots & \mathbf{A}_{0,1N} \\
		\mathbf{A}_{0,21} & \hat{\mathbf{A}}_2 + \mathbf{A}_{0,22} & \cdots & \mathbf{A}_{0,2N} \\
		\vdots & \vdots & \ddots & \vdots \\
		\mathbf{A}_{N0,1} & \mathbf{A}_{0,N2} & \cdots & \hat{\mathbf{A}}_N + \mathbf{A}_{0,NN} \\
	\end{bmatrix}
	\begin{bmatrix}
		\mathbf{u}_1 \\ \mathbf{u}_2 \\ \vdots \\ \mathbf{u}_N
	\end{bmatrix}
	=
	\begin{bmatrix}
		\mathbf{f}_1 \\ \mathbf{f}_2 \\ \vdots \\ \mathbf{f}_N
	\end{bmatrix}
	\label{matrix_assembly}
\end{equation}
where boldface variables denote matrices or vectors for the respective discretized operators. We use the strong form of the boundary integral operators, which is equivalent to mass-matrix preconditioning of the linear system~\cite{betcke2020product}. Notice that all matrix blocks are dense, modeling all-to-all interactions between the mesh elements on the bubble surfaces.

\subsubsection{Computational acceleration}
\label{sec:acceleration}

The BEM's dense matrix arithmetic has a high computational complexity in terms of calculation time and memory footprint. Accelerators like fast multiple methods~\cite{gumerov2005fast} and hierarchical matrix compression~\cite{betcke2017computationally} effectively improve the computational efficiency for large-scale simulations~\cite{jelich2022fast}. However, these techniques require high implementation effort and optimizing algorithmic details to achieve significant speedups. Furthermore, our simulations use relatively coarse meshes and need accurate calculations of cross-interactions between bubbles at many frequencies near the Minnaert resonance. For these reasons, we will not use these approaches but design a custom acceleration scheme for structured bubble arrays that exploits the translation invariance of Green's functions.

We assume all $N$ bubbles to form a two-dimensional array with constant separation distance. Hence, there are $N^2$ interactions between the bubbles, including themselves, as modeled by the Calderón blocks~\eqref{calderon}. Crucially, Green's function is precisely the same when different pairs of bubbles have the same separation distance. Therefore, these Calderón blocks are identical and can be assembled once and reused for all other blocks with the same separation distance. In other words, the number of unique interactions is much smaller than the all-to-all interactions, and the original system~\eqref{PMCHWTFormulation} can be implemented as a matrix-free algorithm that involves unique Calderón blocks only. More precisely, there are $2(N-1)+1$ unique bubble interactions in a rectangular array: one self-interaction and two cross-interactions since the direction matters. Every other interaction between bubbles in the array is a duplicate of these.

We can further reduce computation time by considering the symmetry of interactions between pairs of bubbles. Specifically, $\mathbf{V}_{0,ij} = \left( \mathbf{V}_{0,ji} \right)^T$, $\mathbf{K}_{0,ij} = \left(  \mathbf{T}_{0,ji} \right)^T$,  $\mathbf{T}_{0,ij} = \left( \mathbf{K}_{0,ji} \right)^T$, $\mathbf{D}_{0,ij} = \left( \mathbf{D}_{0,ji} \right)^T$, and $\mathbf{T}_{j} = (\mathbf{K}_{j})^T$; by definition of these operators~\eqref{eq:sl}--\eqref{eq:hs}. Hence, we only assemble the boundary integral operators for $i \leq j$ and transpose them to obtain the operators for $i > j$. Furthermore, the self-interactions require the assembly of three out of four operators.

To summarize, the computational complexity of the standard BEM is $\mathcal{O}(N^2)$, growing quadratically with the bubble count~$N$. Our custom acceleration technique reduces this to $\mathcal{O}(N)$ for calculation time and data storage, but only for structured configurations of bubbles. Specifically, the rectangular configuration allows us to exploit the symmetry and translation-invariance of Green's function. As a final remark, the computational complexity is independent of the separation distance, since the BEM does not mesh the space between the bubbles.

\section{Results}
\label{sec:results}

This section presents numerical results from computational experiments about collective bubble resonances, performed with the mass-spring model and the BEM for the Helmholtz equation. We analyze the behavior of separate modes of increasingly large bubble arrays in Section~\ref{sec:results:modes} and the acoustic field at resonance frequencies in Section~\ref{sec:results:field}.

\subsection{Parameter settings}

All simulations use a speed of sound of 1480~m/s and density of 997~kg/m$^3$ in water, a speed of sound of 343~m/s and density of 1.225~kg/m$^3$ in air, and ambient pressure $P_A=1.01 \cdot 10^5$~Pa. The bubble radius and incident pressure amplitude are normalized to $a=1$ and $P_\mathrm{inc}=1$, respectively.

The BEM uses piecewise linear test and basis functions, defined on a triangular surface mesh whose elements measure at most a sixth of the bubble radius. The iterative GMRES algorithm solves the linear system without restart, a relative termination criterion of $10^{-5}$, and mass-matrix preconditioning. The discrete BEM matrices are constructed using the open-source Python library bempp-cl~\cite{betcke2021bempp}, with multithreading performed through the PyOpenCL package. The software was previously validated against coupled spherical harmonics for bubble pairs~\cite{wout2021proximity} and various numerical methods for ultrasonics~\cite{aubry2022benchmark}. Our custom acceleration for rectangular arrays, explained in Section~\ref{sec:acceleration}, is made publicly available on GitHub\footnote{R.~Jerez~Boudesseul and E.~van~'t~Wout: \url{www.github.com/rudyjb24/Fast-BEM}}. The computational simulations were performed on a 12-core Intel(R) Xeon E5-2658A v3, 32~GB RAM machine.

\subsection{Modal analysis of collective resonances}
\label{sec:results:modes}

The mass-spring model for bubble dynamics, presented in Section~\ref{sec:formulation:mass_spring}, is a system of ordinary differential equations that can be decoupled into independent modes~\eqref{eq:mode:def}. The characteristics of each resonance mode, such as frequency and quality factor~$Q$, depend on the eigenvalues of Green's matrix~\eqref{eq:greenmatrix:spectrum}. The corresponding eigenvectors determine the excitation factor~\eqref{eq:excitationfactor} of each resonance mode. The spectrum, that is, the collection of all eigenvalues, of Green's matrix is interesting in itself~\cite{rusek2000random, goetschy2011non, skipetrov2016finite} and has implications for different simulations of acoustic metamaterials~\cite{leroy2005bubble}, condensed matter physics~\cite{zeravcic2011collective}, and quantum mechanics~\cite{gaspard2022resonance}. Hence, Section~\ref{sec:results:modes:eigenvalues} shows the spectrum for increasingly large collections of point scatterers. Section~\ref{sec:results:modes:resonances} applies these results to collective resonances of air bubbles immersed in water, showing an initial downshift and subsequent upshift in the fundamental resonance frequency of large bubble arrays. All results correspond to the homogeneous problem where Green's matrix is evaluated at the Minnaert frequency of a single bubble.

\subsubsection{Eigenvalues of Green's matrix}
\label{sec:results:modes:eigenvalues}

For a pair of bubbles with a distance~$d$ between their centers, the eigenvalues of the Green's matrix are given by $\pm e^{\imath k d}/(4\pi d)$, and the corresponding eigenvectors are $[ 1, \, \pm 1 ]$. These are the well-known symmetric and anti-symmetric proximity resonances~\cite{wout2021proximity} and, when visualized for different distances, depict intertwined spirals in the complex plane~\cite{gaspard2022resonance}. For three bubbles in an equilateral triangular configuration, with distance~$d$ between them, the positive eigenvalue $e^{\imath k d}/(2\pi d)$ has a constant eigenvector $[1, \, 1, \, 1]$, while the negative eigenvalue $-e^{\imath k d}/(4\pi d)$ is degenerate with eigenspace spanned by the vectors $[1, \, 0, \, -1]$ and $[1, \, -2, \, 1]$, which is consistent with known superresonances~\cite{feuillade1995scattering, illesinghe2009eigenmodal}. Green's matrix generally is non-normal with complex-valued eigenvectors that do not include the vector of equal entries. One such example is three bubbles located in a line. Furthermore, the eigendecomposition needs to be calculated numerically for any but trivial configurations.

\begin{figure}[!ht]
	\centering
	\includegraphics[width=\columnwidth]{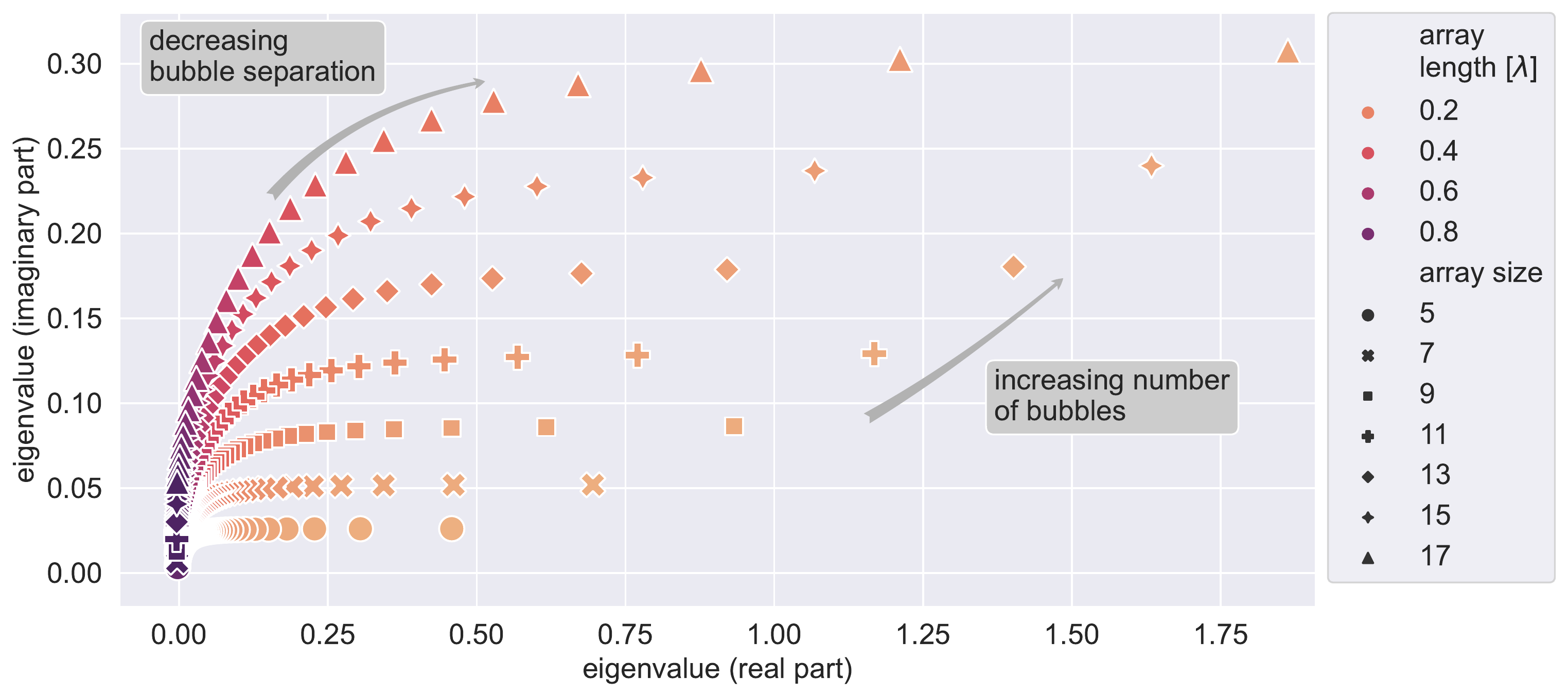}
	\caption{Each marker corresponds to the eigenvalue of the fundamental mode of Green's matrix for a different array configuration. The separation distances are $d_m=ma$ for $m=2,3,4,\dots$ while $(n-1)d_m < \lambda$, that is, the array side covers less than a wavelength. The color indicates the array side length, normalized by the wavelength at Minnaert resonance. The marker type indicates the bubble count in the $n \times n$ square array.}
	\label{fig:spectrum:smallarrays}
\end{figure}

The dominant response of a bubble array at resonance is the fundamental mode corresponding to the largest eigenvalue of Green's matrix. Figure~\ref{fig:spectrum:smallarrays} shows the influence of the array configuration on the fundamental mode where bubbles oscillate in the same phase. The results show that in the cases of: (1) an increasing number of bubbles at fixed bubble separation; and (2) a decreasing bubble separation at fixed bubble count, the largest eigenvalue of the configurations grows in both its real and imaginary part, albeit in different ways. The top right marker corresponds to the configuration of 17~bubbles located at a distance of two bubble radii from each other, covering less than 10\% of a wavelength in total.

\begin{figure}[!ht]
	\centering
	\includegraphics[width=\columnwidth]{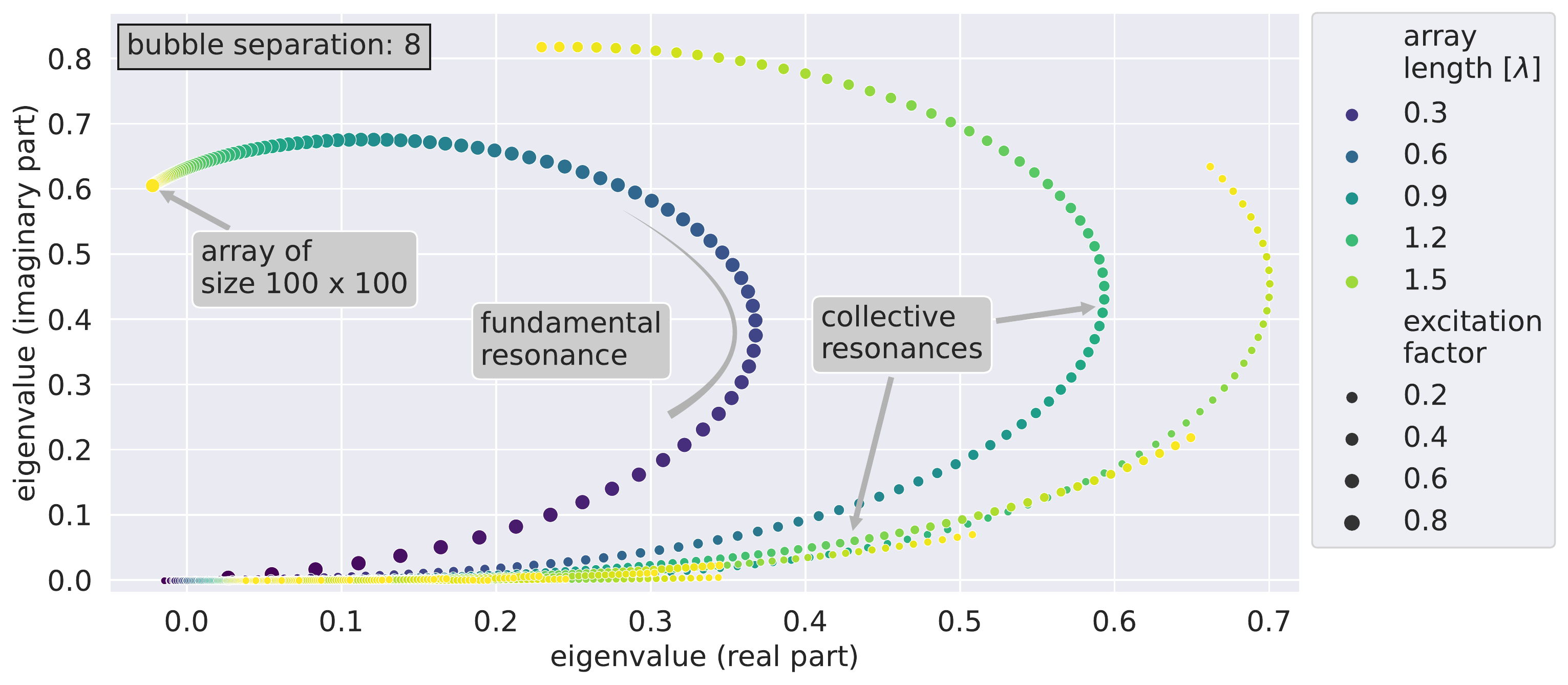}
	\includegraphics[width=\columnwidth]{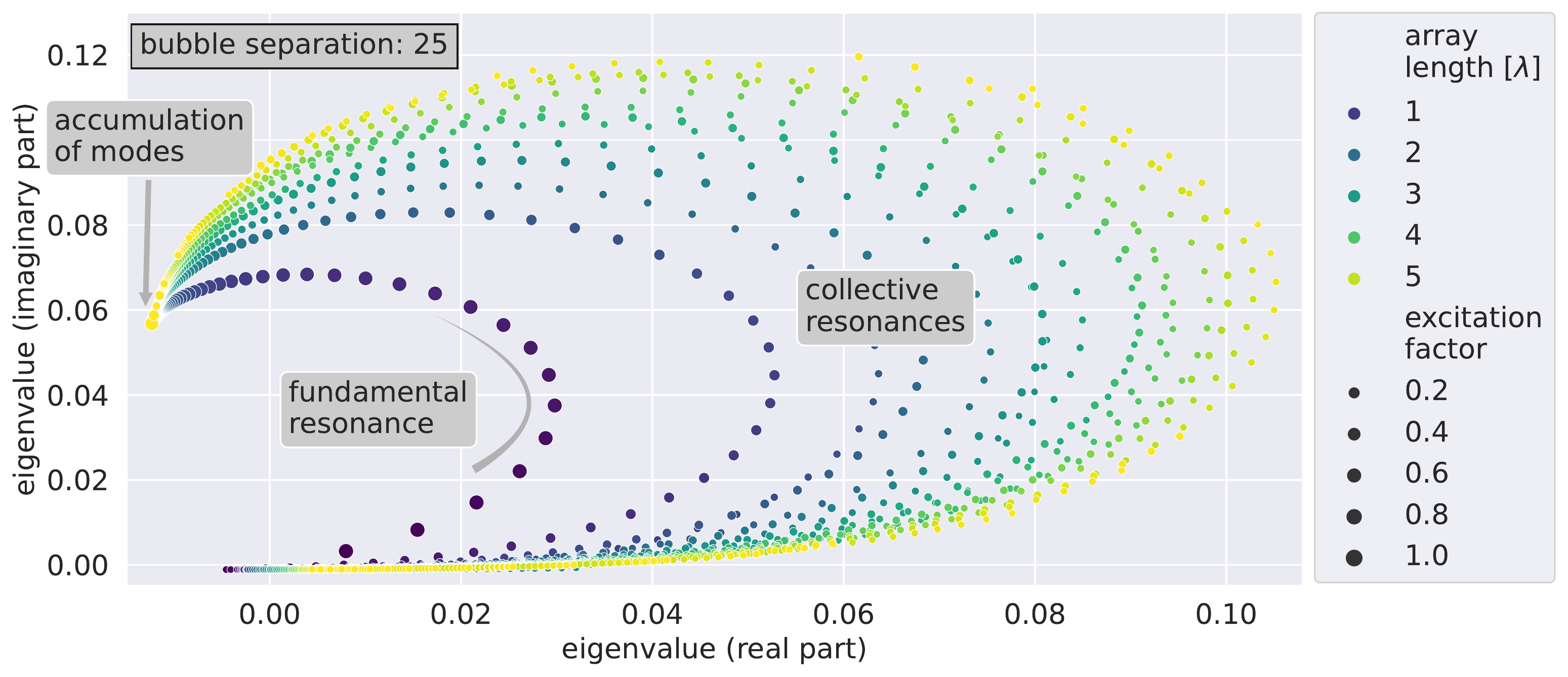}
	\caption{The eigenvalues of Green's matrix for the dominant resonances, that is, the modes with an excitation factor larger than 1\%. The marker color indicates the side length of the $n \times n$ square bubble array for $n=2,3,4,\dots,100$, and the marker size is proportional to the excitation factor.}
	\label{fig:spectrum:largearrays}
\end{figure}

When a bubble array covers more than a wavelength, there are increasing opportunities for bubbles to oscillate with opposite phases. This influences the modes of the mass-spring model, with superresonances gaining dominance. Let us consider a square array with constant separation distance and an increasing number of bubbles to investigate this phenomenon. Figure~\ref{fig:spectrum:largearrays} shows the eigenvalues of Green's matrix for each configuration, but only for modes that are excited significantly by the incident plane wave. Specifically, the eigenvalues corresponding to superresonances with an excitation factor smaller than 1\% are not shown, in order to avoid clutter. The results show that configurations extending less than half a wavelength have one dominant mode with an eigenvalue increasing with bubble count. After this, the dominant eigenvalue reverses and approaches a limit with a negative real value and a positive imaginary value. Meanwhile, other modes appear at large array sizes, with a significant excitation factor and a large eigenvalue. These are interpreted as collective resonances, since they only occur when the array covers more than half a wavelength. These collective modes accumulate at the same value in the complex plane as the fundamental resonance when the array covers several wavelengths.

\subsubsection{Collective resonances in mass-spring systems}
\label{sec:results:modes:resonances}

The nonlinear relations~\eqref{eq:tilde:omega} and~\eqref{eq:tilde:q} determine the resonance frequency and quality factor of each mode from the eigenvalues of Green's matrix. Generally, a larger real part for the eigenvalue indicates a larger downshift in resonance frequency, and a larger imaginary part indicates a broader peak with a smaller quality factor. Hence, the curves in Figure~\ref{fig:spectrum:largearrays} suggest an initial downward and subsequent upward shift in resonance frequency, with an overall widening of the peak for increasingly large bubble arrays.

\begin{figure}[!ht]
	\centering
	\includegraphics[width=\columnwidth]{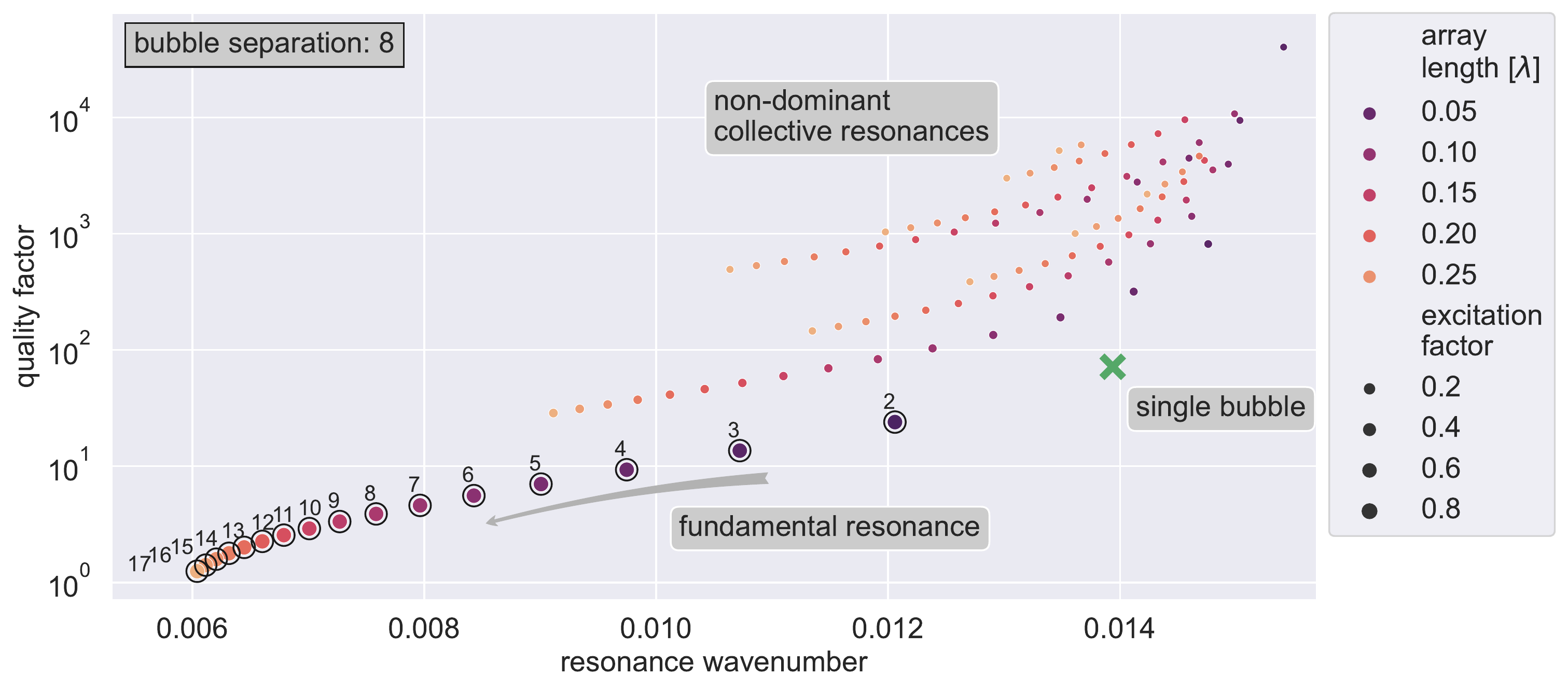}
	\includegraphics[width=\columnwidth]{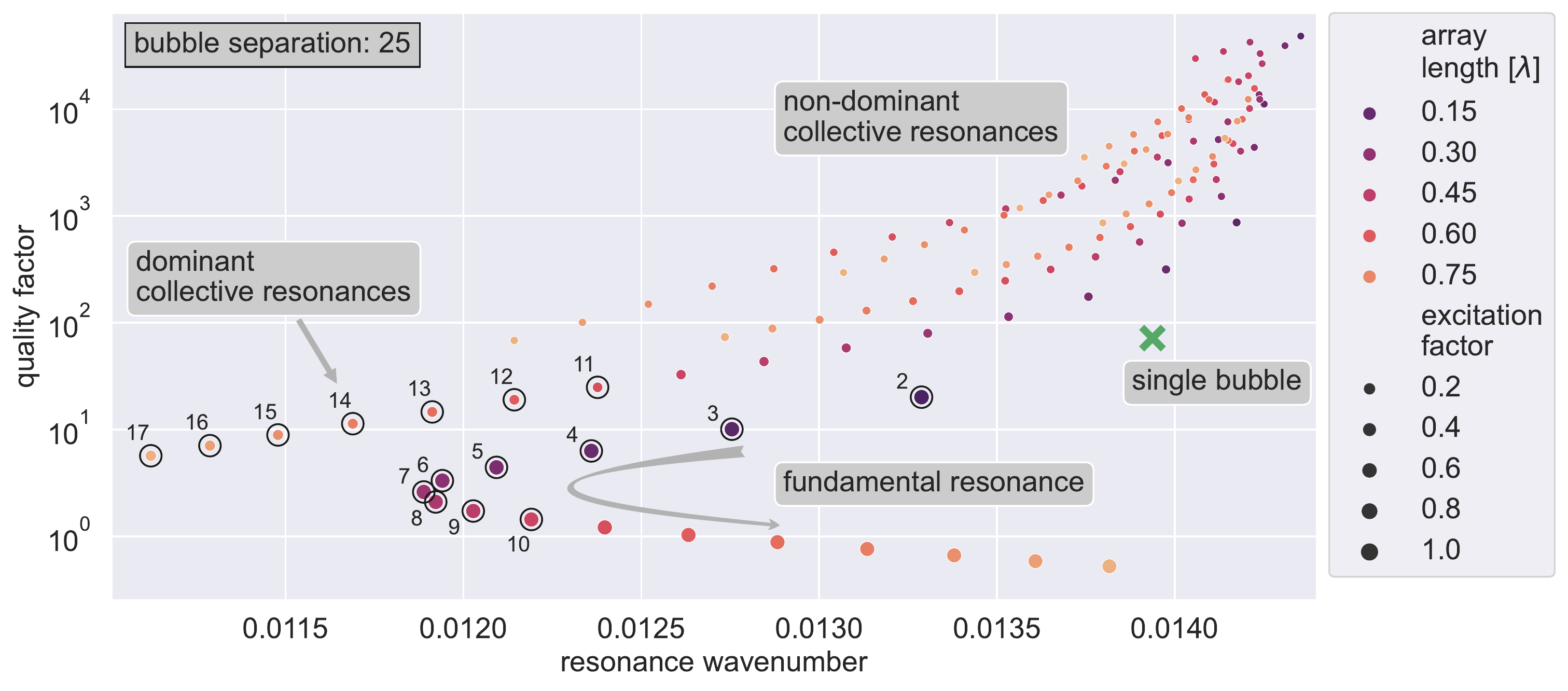}
	\caption{The frequency and quality factor~$Q$ of the dominant resonances according to the modal analysis of the mass-spring system. Specifically, only the modes with an excitation factor larger than 1\% are shown. The marker color indicates the side length of the $n \times n$ square bubble array for $n=2,3,4,\dots,17$, and the marker size is proportional to the excitation factor. The numbers highlight the mode with the smallest resonance frequency for the specific value of $n$.}
	\label{fig:spectrum:resonances}
\end{figure}

Figure~\ref{fig:spectrum:resonances} shows the modes' resonance wavenumber and quality factor~$Q$, calculated from the eigenvalues of Green's matrix. The configurations are square bubble arrays with fixed separation distance and increasing size. The superresonances that are not excited by a perpendicular plane wave were filtered out, and the single bubble was included as a reference. For a short separation distance of eight radii, increasing the number of bubbles in the array causes a downshift in frequency and broadening of the fundamental resonance. The collective resonances are never dominant: their excitation factor is small, and their frequency shifts remain limited. In contrast, considering a separation distance of 25~radii, the fundamental resonance indicates a return in frequency towards higher values when the bubble array reaches about half the wavelength. Moreover, when more than a hundred ($10 \times 10$) bubbles are present, the second collective resonance mode becomes dominant: it causes the largest downshift in frequency, even though it still has a smaller excitation factor than the fundamental resonance mode. Hence, we interpret the \emph{fundamental} mode as the mode with bubbles oscillating in the same phase, not necessarily the mode that causes the largest frequency shift.

\subsubsection{Discussion}

The modal analysis of the mass-spring model confirms important phenomena of collective bubble resonances observed in the literature. For example, the fundamental resonance of small bubble ensembles displays a downshift in frequency and a widening of the response peak for increasing bubble count~\cite{leroy2005bubble, roshid2020extraction} and shorter separation distances~\cite{payne2005symmetric, illesinghe2009eigenmodal}. Furthermore, collective resonance modes accumulate at a specific eigenvalue with a negative real part when increasing the bubble count. Assuming this phenomenon continues for $N \to \infty$, these results are consistent with the upward shift in resonance frequency for screens~\cite{leroy2009transmission, calvo2015underwater, leroy2018acoustics}.

Figures~\ref{fig:spectrum:largearrays} and~\ref{fig:spectrum:resonances} show a changing nature of the collective resonances when the size of a bubble array reaches half the wavelength. The fundamental resonance remains the mode most strongly excited by a perpendicular plane wave. However, its peak continues to widen while the earlier downshift in resonance frequency reverses into an upshift compared to a single bubble. Furthermore, higher-order collective resonance modes become significant in two ways: (1) they are excited strongly by the perpendicular forcing; and (2) they display a large downshift in resonance frequency.

\subsection{Acoustic field at collective resonances}
\label{sec:results:field}

\begin{figure}[!ht]
	\centering
	\includegraphics[width=\columnwidth]{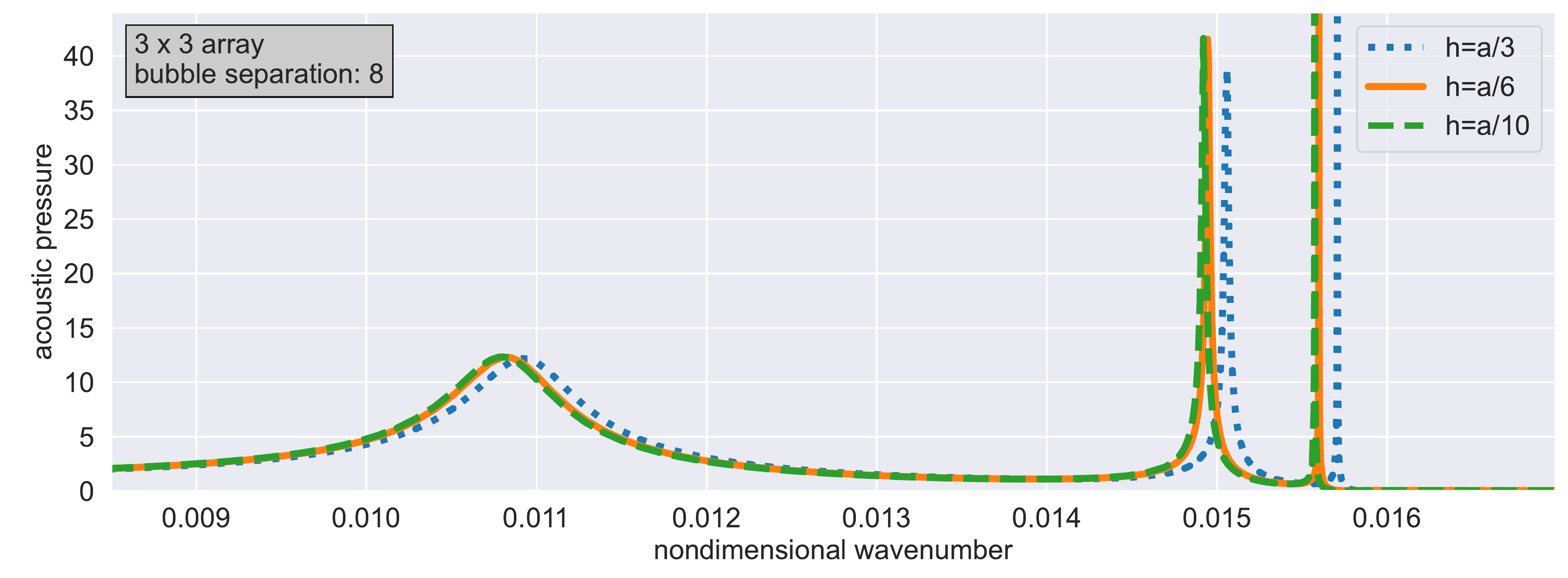}
	\caption{The lines indicate the acoustic pressure at location $(3a,0,0)$, using the BEM with mesh width~$h$. The bubbles in the $3 \times 3$ array have radius $a$ and are separated by a distance of $8a$.}
	\label{fig:refinement}
\end{figure}

The modal analysis gives valuable information about separate modes of collective oscillations in the bubble array. However, the underlying mass-spring model uses strong assumptions, principally treating each bubble as a monopole point scatterer. While reasonable at the low frequencies considered here, this assumption has significant limitations in cases such as short separation distances~\cite{leroy2018acoustics, wout2021proximity}. Furthermore, physical experiments measure the total acoustic pressure, not separate modes. Alternatively, the Helmholtz equation~\eqref{eq:helmholtz} accurately models the acoustic field scattered by the bubbles. This system of partial differential equations has an analytical solution for a single spherically-shaped bubble~\cite{feuillade2012superspheroidal}, and numerical methods efficiently approximate the acoustic field from interacting bubbles. The BEM with our custom acceleration algorithm, explained in Section~\ref{sec:formulation:bem}, has linear computational complexity with the number of bubbles in a rectangular array and is, therefore, one of the most efficient numerical methods for studying acoustic bubble dynamics. We perform multiple BEM simulations at hundreds of frequencies to analyze the resonance behavior of the bubble arrays. In contrast, the modal analysis of the mass-spring system always evaluates Green's matrix at the Minnaert resonance frequency of a single uncoupled bubble. The BEM's grid dependency converged for a mesh size of $a/6$, as confirmed by the computational results in Figure~\ref{fig:refinement}.

\subsubsection{Scattering amplitude of small ensembles}

\begin{figure}[!ht]
	\centering
	\includegraphics[width=\columnwidth]{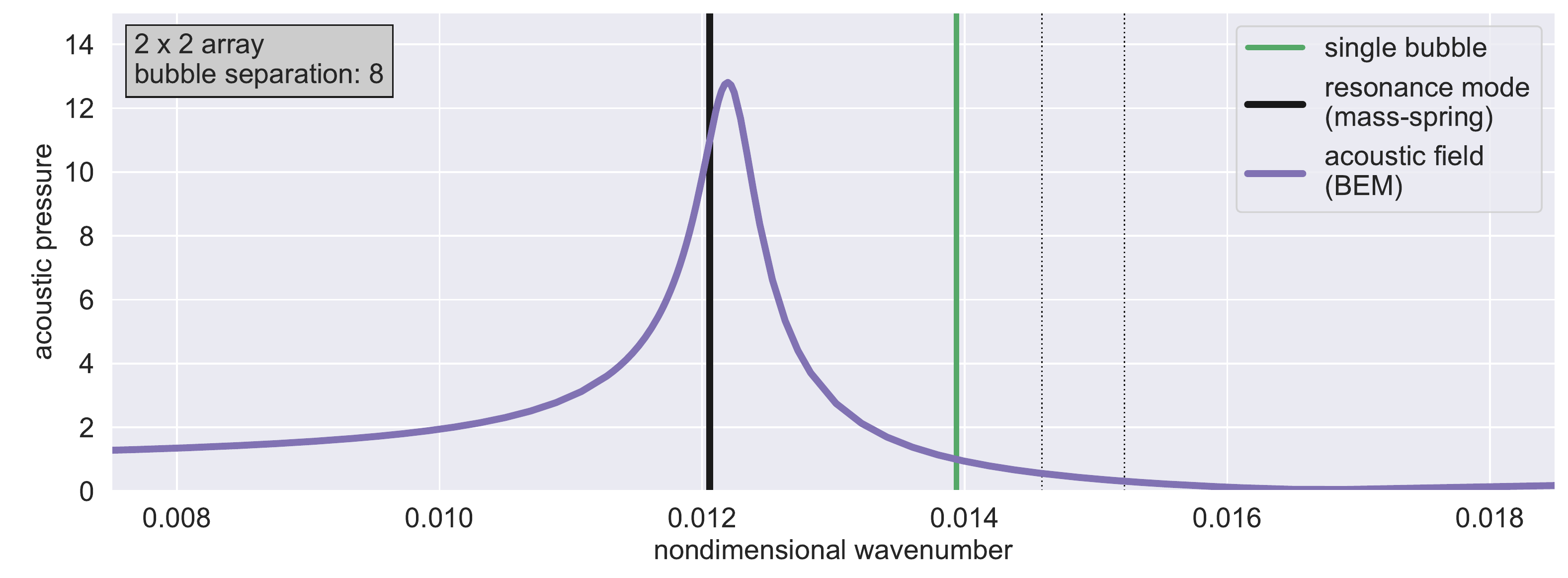}
	\includegraphics[width=\columnwidth]{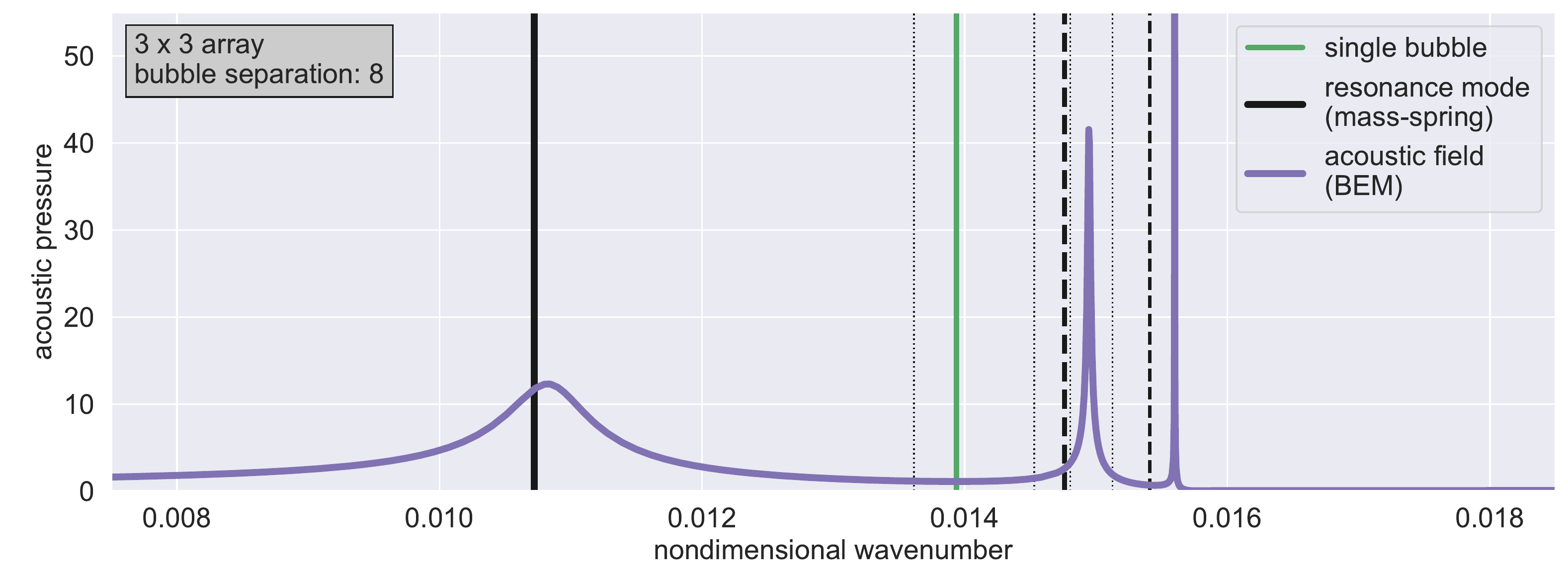}
	\caption{The purple curve indicates the acoustic pressure at location $(3a,0,0)$, using the BEM. The incident plane wave field propagates in positive $x$-direction, and the array is in the $(y,z)$-plane, centered at the global origin. The vertical lines indicate the resonance frequencies of a single bubble (green) and the array (black), using the mass-spring model. Their thickness indicates the excitation factor.}
	\label{fig:amplitude:smallarray}
\end{figure}

The eigenvalues of Green's matrix for a square array with four bubbles and separation distance~$d$ are $\sigma_0 = e^{-\imath k d}/(2\pi d) + e^{-2\imath k d}/(8\pi d)$, $\sigma_1 = -e^{-\imath k d}/(2\pi d) + e^{-2\imath k d}/(8\pi d)$, and the degenerate $\sigma_2 = -e^{-2\imath k d}/(8\pi d)$. The eigenvector of the fundamental mode has identical entries, indicating all the bubbles oscillate in phase. The other eigenvectors represent collective motions with neighboring bubbles oscillating in opposite phases. Hence, only the fundamental mode is excited by a perpendicular plane wave, and its resonance frequency is lower than for a single bubble. Figure~\ref{fig:amplitude:smallarray} shows the modes' frequencies together with the acoustic pressure variation at a distance of three bubble radii from the array, in the forward direction. Clearly, the peak in acoustic pressure calculated by the BEM corresponds to the fundamental resonance at $k a = 0.01206$ of the mass-spring model. The small frequency discrepancy observed is due to the different modeling approaches. The field scattered from a square array with $3 \times 3$ bubbles, shown in Figure~\ref{fig:amplitude:smallarray}, also indicates a broadening and downshift of the fundamental resonance. The two sharp peaks in the scattered field are higher-order collective modes with excitation factors 0.095 and 0.013, respectively, at frequencies indicated by the dashed vertical lines. The four dotted vertical lines indicate the frequencies of the six superresonances---two of which are degenerate---that are not excited by the perpendicular forcing. Subsequent figures will omit these unexcited superresonances.

\subsubsection{Scattering amplitude of large arrays}

\begin{figure}[!ht]
	\centering
	\includegraphics[width=\columnwidth]{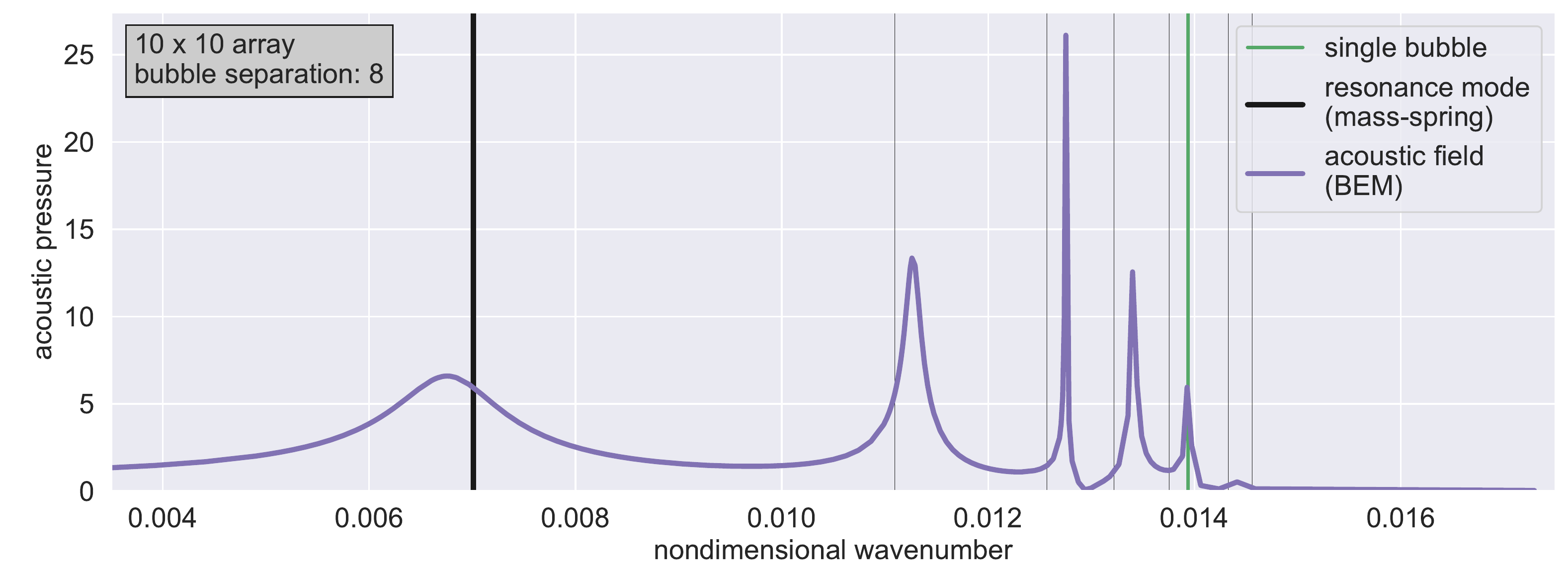}
	\includegraphics[width=\columnwidth]{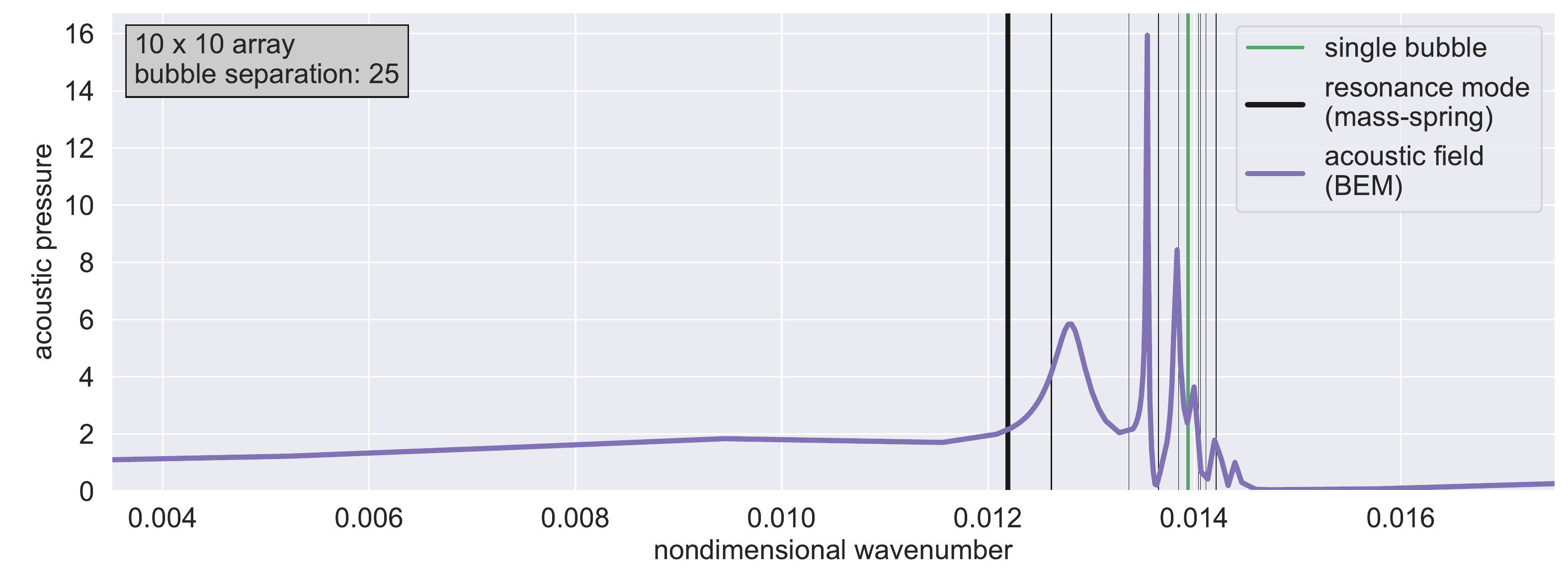}
	\caption{The purple curve indicates the acoustic pressure at location $(3a,0,0)$, using the BEM. The incident plane wave field propagates in positive $x$-direction, and the array is in the $(y,z)$-plane, centered at the global origin. The vertical lines indicate the resonance frequencies of a single bubble (green) and the array (black), using the mass-spring model. Their thickness indicates the excitation factor.}
	\label{fig:amplitude:largearray}
\end{figure}

The incident plane wave excites an increasing number of modes when arrays contain more bubbles. For example, Figure~\ref{fig:amplitude:largearray} shows the amplitude of the scattered field at a distance of three bubble radii in the forward direction for arrays with $10 \times 10$ bubbles. The many sharp peaks are collective resonances, some of them with a frequency lower, others with a frequency higher than a single bubble. Notably, the fundamental resonance is still visible as a broad peak around the mode's frequency when the separation distance is 8~radii. In contrast, the same $10 \times 10$ array with a separation distance of 25~radii covers more than half the wavelength, and the fundamental and collective resonances can no longer be separated from their acoustic responses due to the broad peaks and similar frequencies.

\begin{figure}[!ht]
	\centering
	\includegraphics[width=\columnwidth]{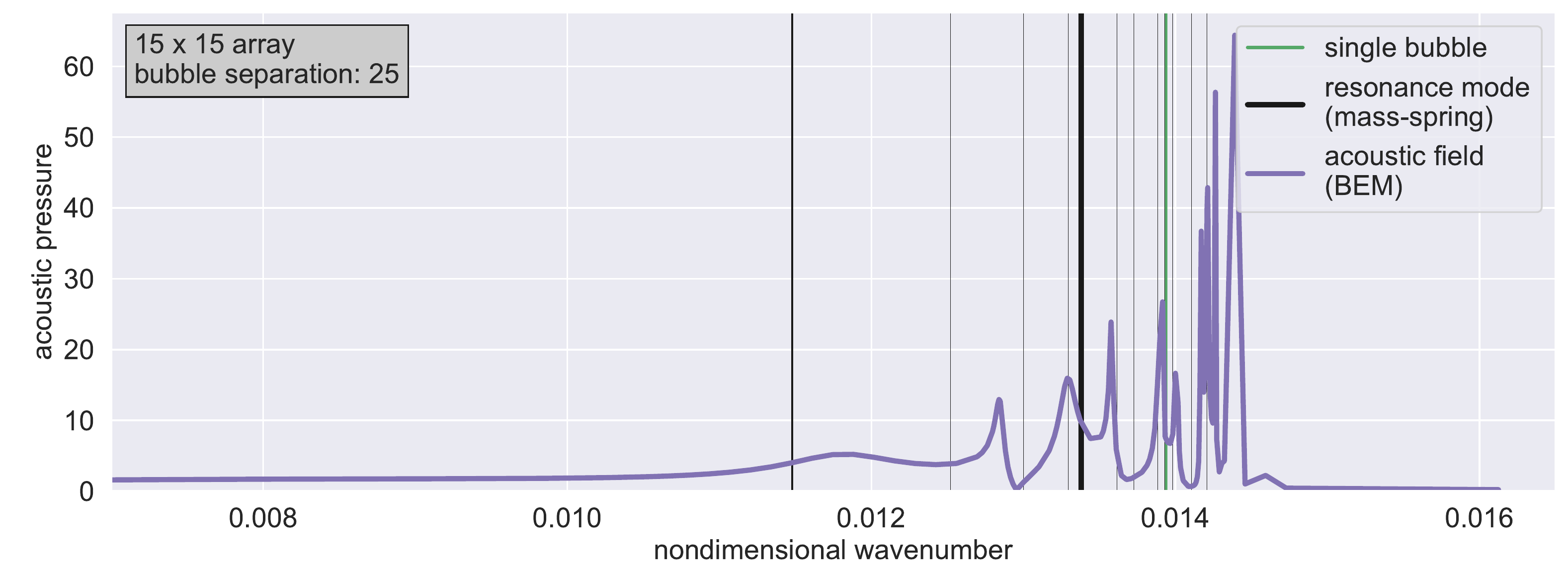}
	\caption{The purple curve indicates the acoustic pressure at location $(3a,0,0)$, using the BEM. The incident plane wave field propagates in positive $x$-direction, and the array is in the $(y,z)$-plane, centered at the global origin. The vertical lines indicate the resonance frequencies of a single bubble (green) and the array (black), using the mass-spring model. Their thickness indicates the excitation factor.}
	\label{fig:amplitude:largestarray}
\end{figure}

Figure~\ref{fig:amplitude:largestarray} shows the acoustic field scattered from a square array of $15 \times 15$ bubbles with a separation distance of 25~radii. A coarser mesh with an element size of $a/3$ was used to limit computation time. The array spans one wavelength across its diagonal, and the fundamental mode can no longer be identified within the overall acoustic field. It is surrounded by collective modes, and its quality factor is so low that the peak amplitude is insignificant.

\subsubsection{Spatial characteristics of resonance modes}

The BEM is a full-wave solver that solves the Helmholtz equation at the surface of each bubble and accurately calculates the acoustic pressure field in the entire three-dimensional space from Eq.~\eqref{eq:bie:representation}. The following results show the acoustic field scattered from an array of $10 \times 10$ bubbles with a separation distance of eight radii at $ka=0.006755$ and $ka=0.011261$. These frequencies correspond to the fundamental resonance and the first high-order collective mode, respectively, visible as peaks in Figure~\ref{fig:amplitude:largearray}. Its characteristics are also seen in a $10 \times 10$ array with a separation distance of 25~radii at frequencies $ka=0.0094$ and $ka=0.0128$, but this case is not shown here for brevity.

\begin{figure}[!ht]
	\centering
	\parbox[h]{.9\columnwidth}{\footnotesize (a) {The fundamental resonance at $ka=0.006755$.}}
	\includegraphics[height=5.7cm]{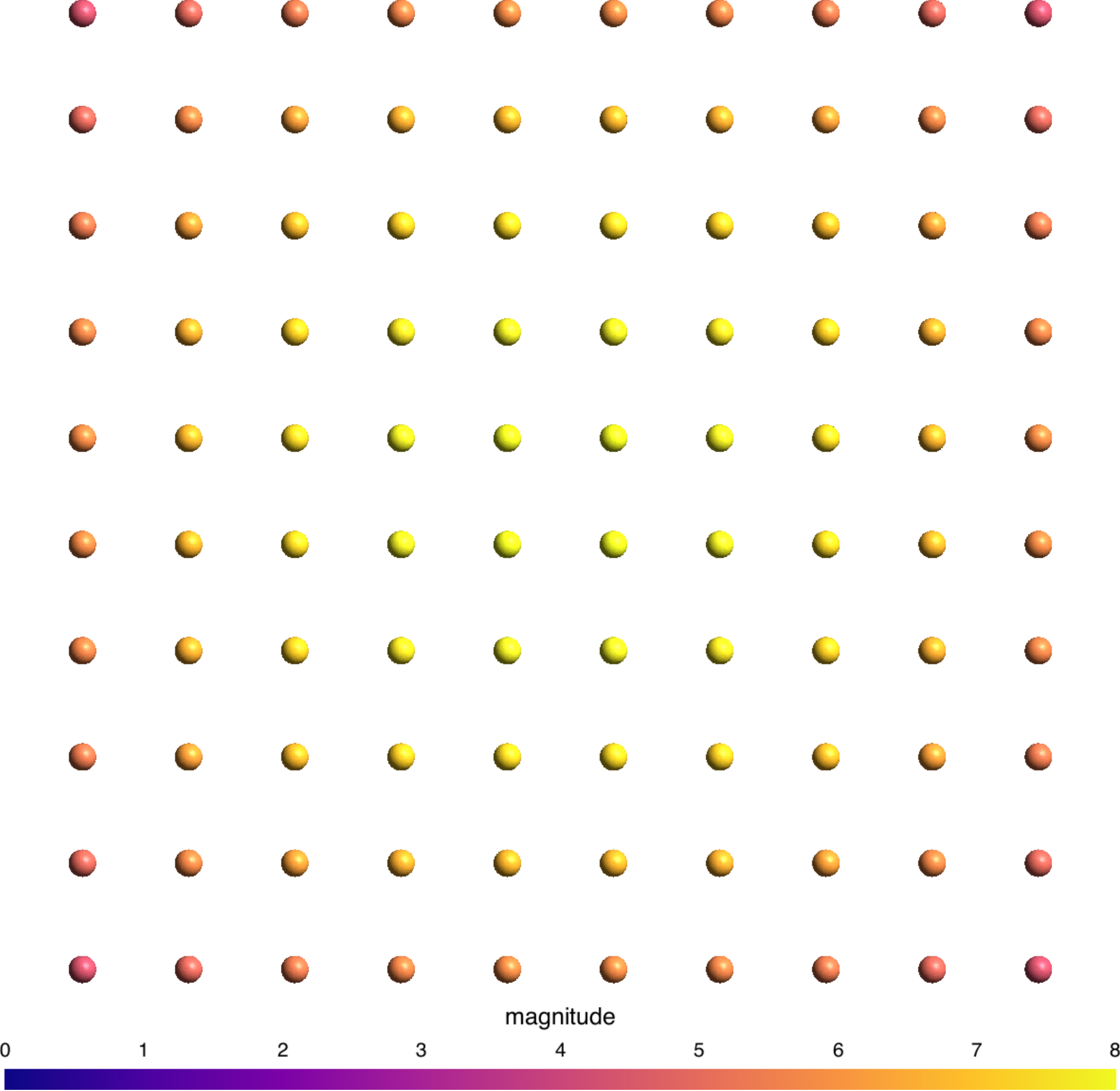}
	\includegraphics[height=5.7cm]{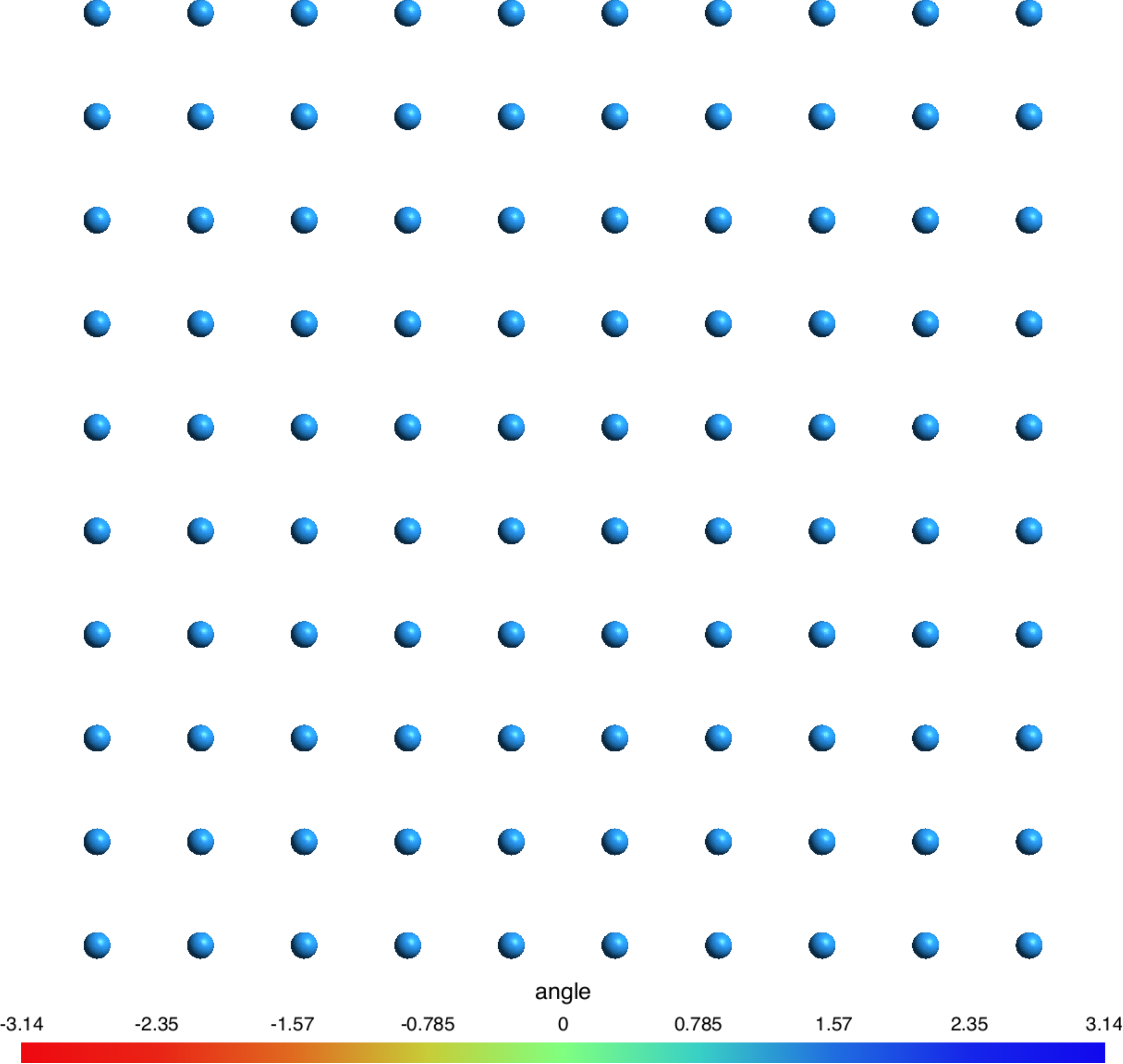}
	\parbox[h]{.9\columnwidth}{\footnotesize (b) {The first high-order collective resonance at $ka=0.011261$.}}
	\includegraphics[height=5.7cm]{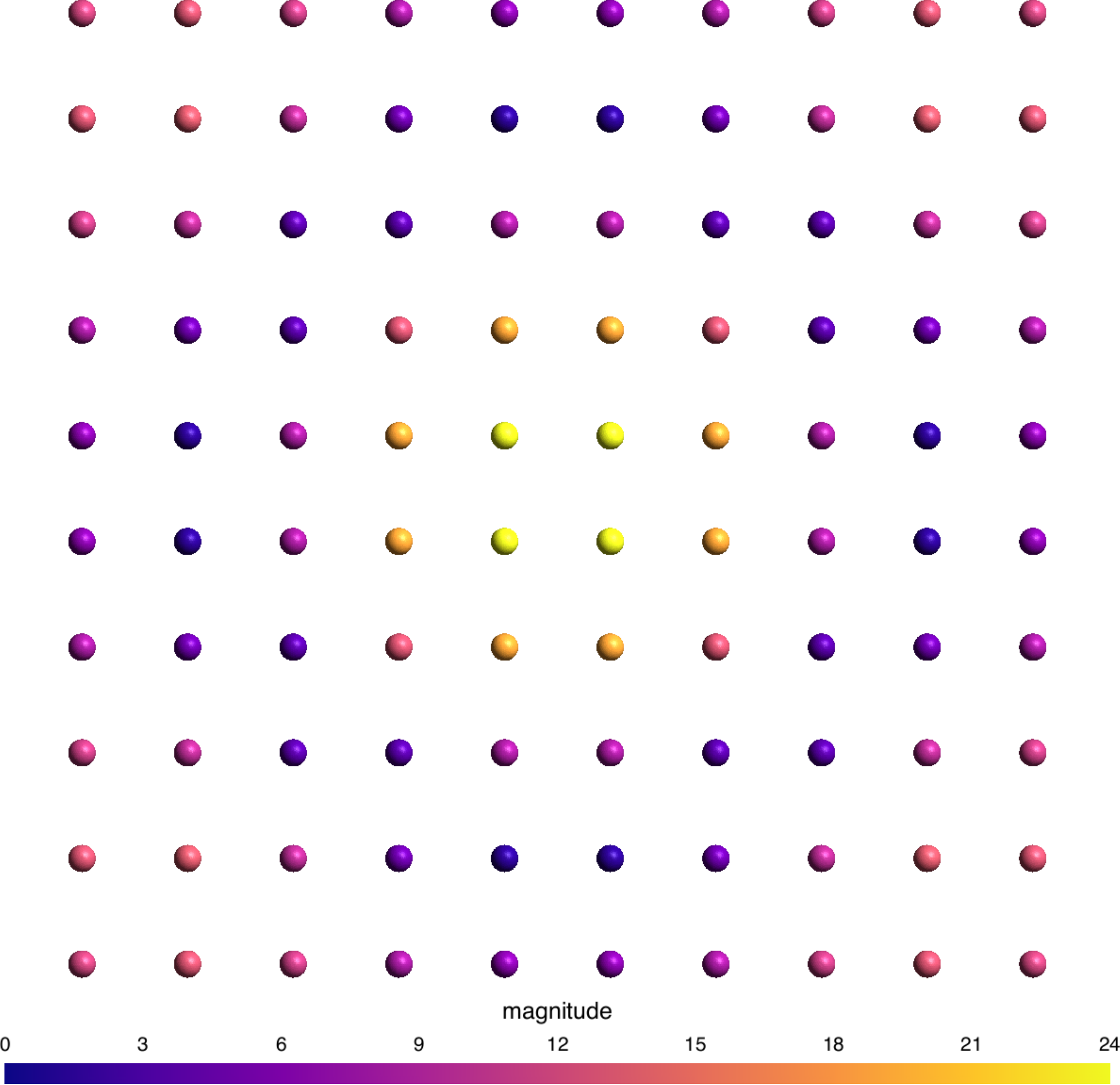}
	\includegraphics[height=5.7cm]{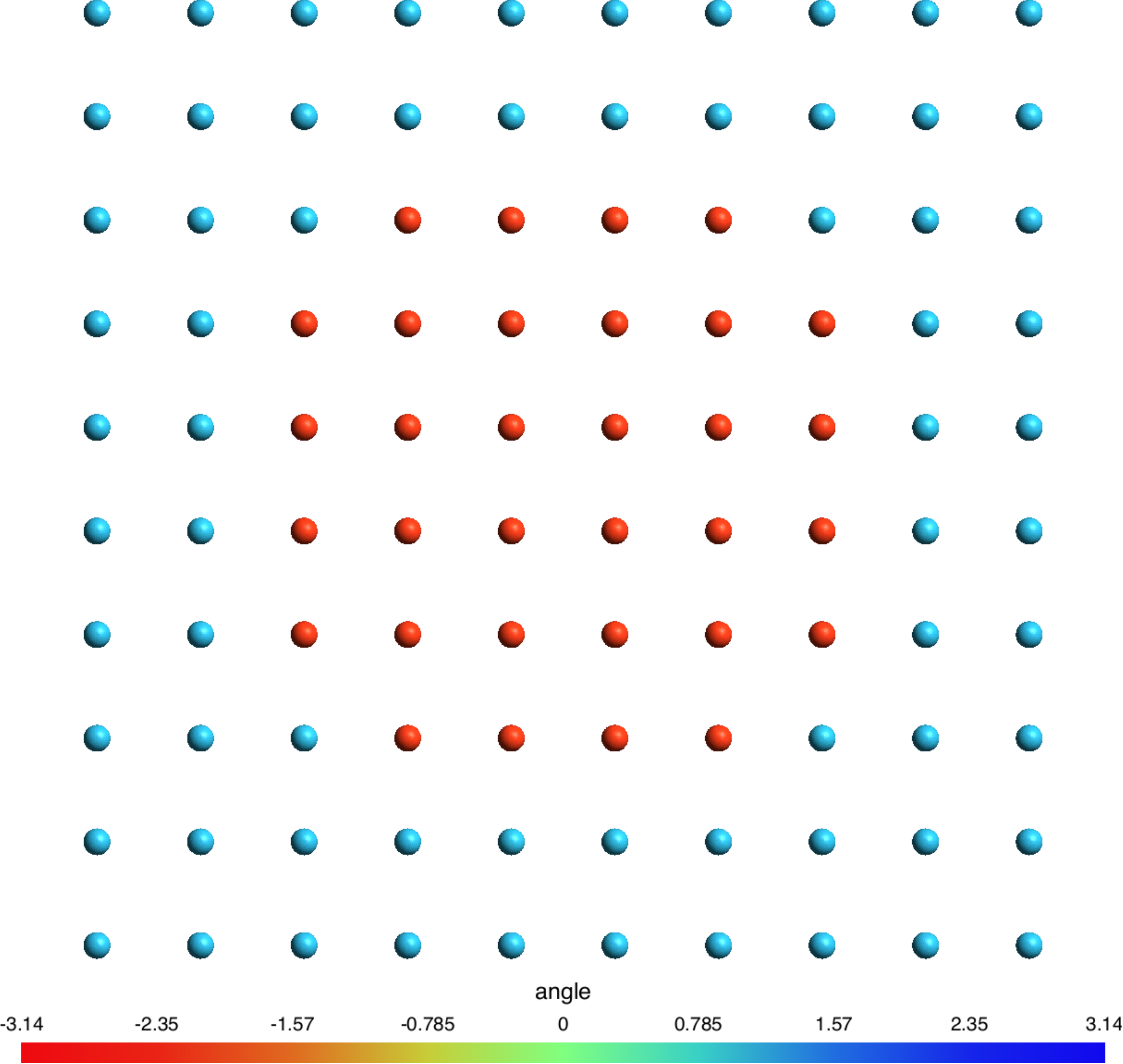}
	\caption{The magnitude (left) and phase (right) of the acoustic pressure at the surface of the bubbles. The array has $10 \times 10$ bubbles with a separation distance of eight radii.}
	\label{fig:surfpot}
\end{figure}

\paragraph{Surface potentials at resonance}
The solution of the BEM includes surface potentials that represent the acoustic field and its normal derivative at each bubble surface. Figure~\ref{fig:surfpot} displays the polar decomposition of the complex-valued acoustic pressure at the bubble surfaces. In the fundamental mode, the center bubbles oscillate more strongly than the bubbles at the array edges, because of the multiple interactions between the bubbles. All bubbles oscillate in the same phase, indicating a monopole resonance. In contrast, the collective resonance mode has center bubbles oscillating in opposite phases to the bubbles at the edges. The bubbles close to others oscillating in opposite phases have a small amplitude, while bubbles oscillate more strongly when surrounded by bubbles with the same phase.

\begin{figure}[!ht]
	\centering
	\parbox[h]{.9\columnwidth}{\footnotesize (a) {The fundamental resonance at $ka=0.006755$.}}
	\includegraphics[height=5.0cm]{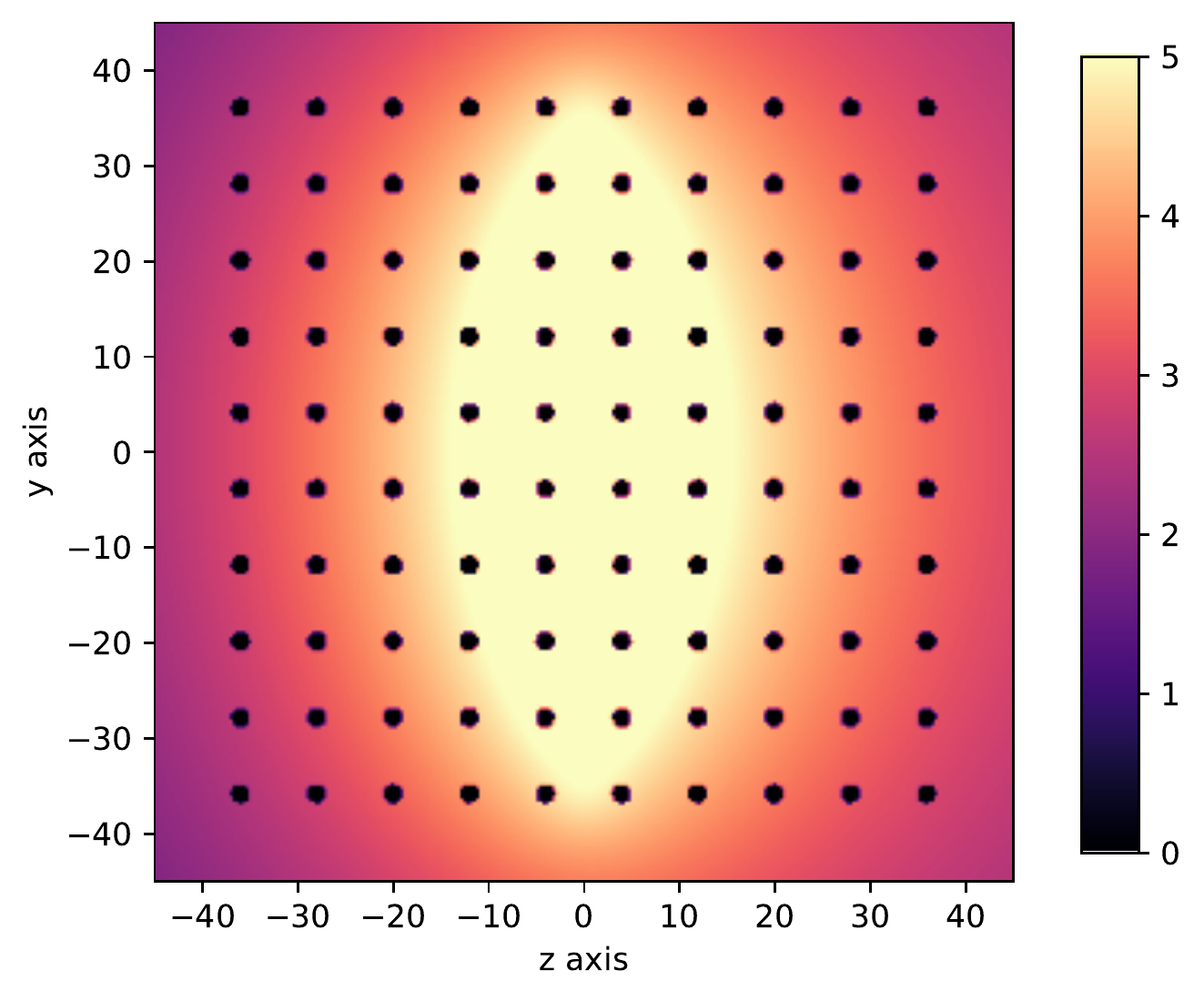}
	\includegraphics[height=5.0cm]{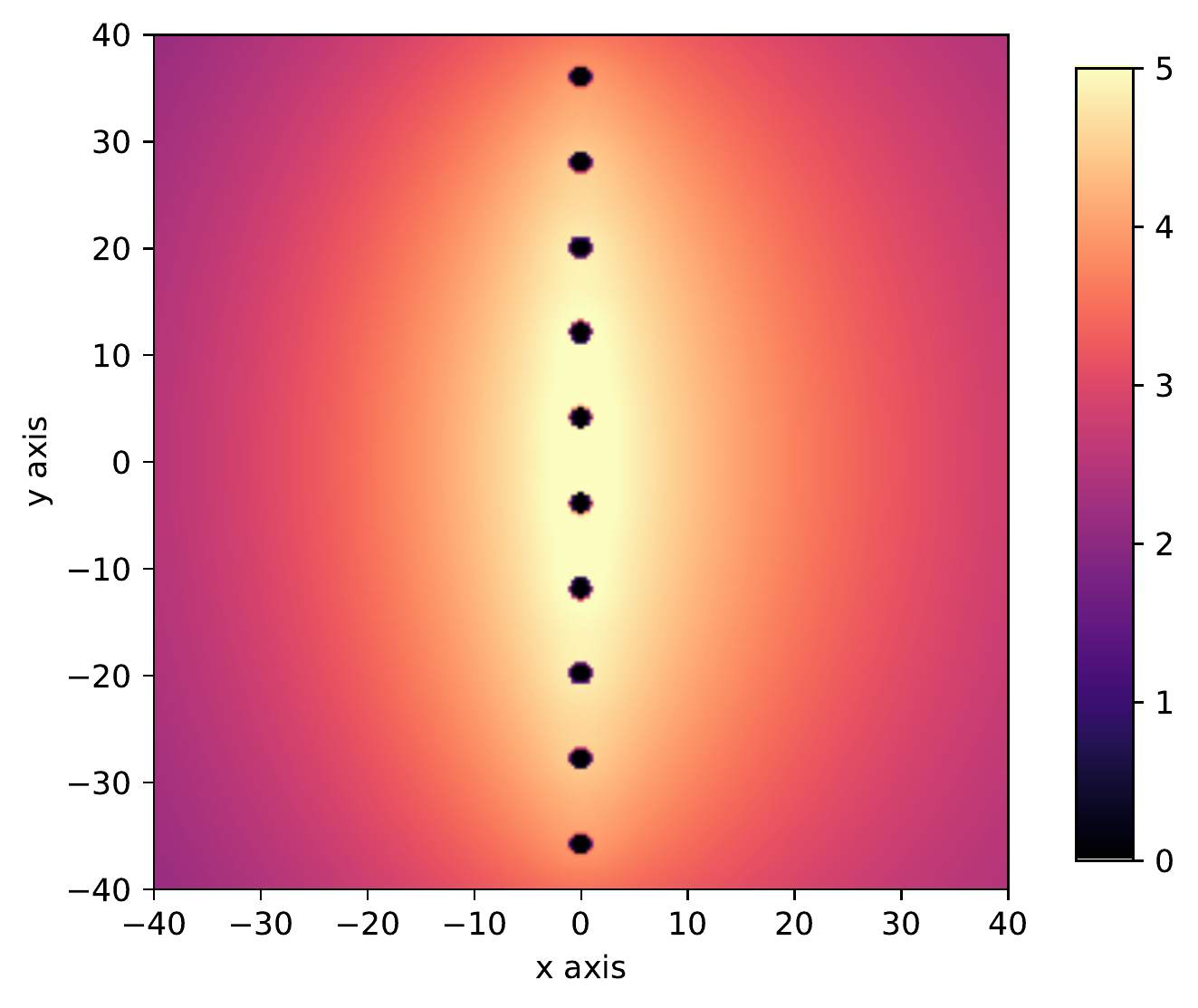}
	\parbox[h]{.9\columnwidth}{\footnotesize (b) {The first high-order collective resonance at $ka=0.011261$.}}
	\includegraphics[height=4.9cm]{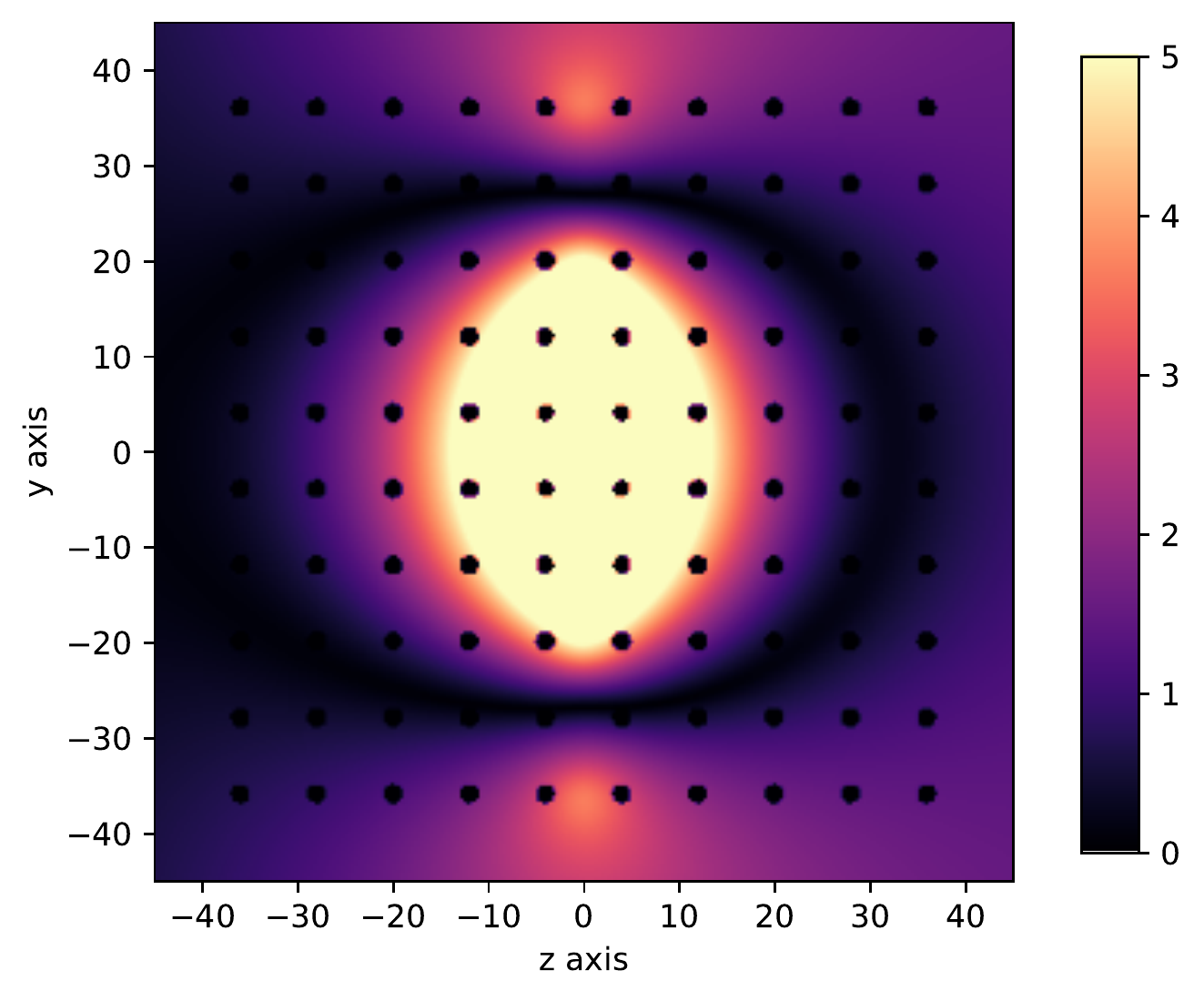}
	\includegraphics[height=4.9cm]{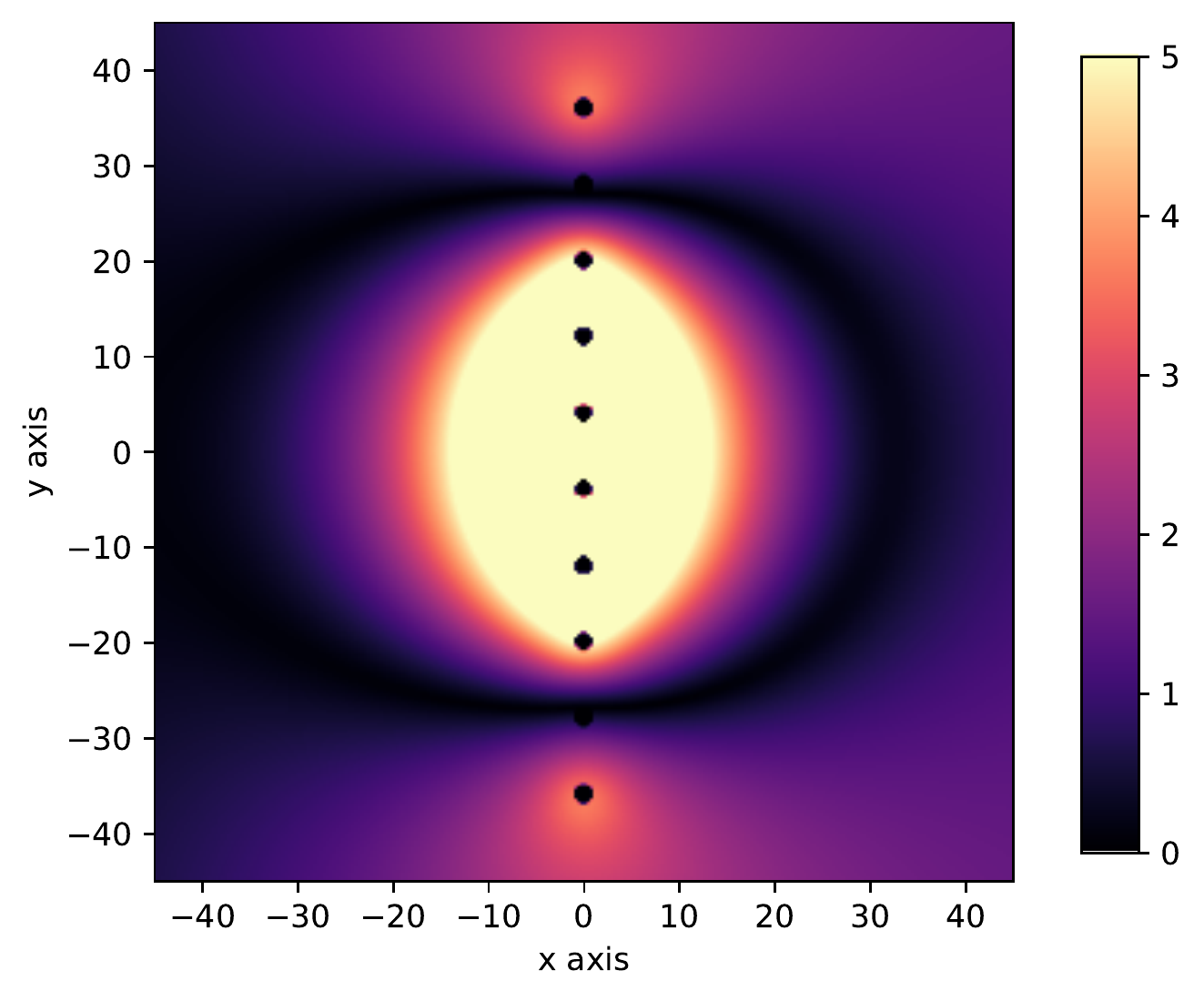}
	\caption{The amplitude of the acoustic pressure at the slices $x=0$ (left) and $z=0$ (right). The array has $10 \times 10$ bubbles with a separation distance of eight radii. The axes are normalized by the bubble radius.}
	\label{fig:field:slice}
\end{figure}

\paragraph{Near field at resonance}
The oscillations of the bubbles generate an acoustic pressure field in the volume around the array. Figure~\ref{fig:field:slice} shows the acoustic pressure amplitude along two slices, one through the array and one perpendicular to it. The field is spherically symmetric at the fundamental resonance, similar to a monopole field. In contrast, the collective resonance displays a shape of a multipole field, with a big hotspot in the middle, and two more minor spots at the far ends. The symmetries in the field are imperfect, due to the array's finite-size effects and influences from other modes.

\begin{figure}[!ht]
	\centering
	\includegraphics[width=.49\columnwidth]{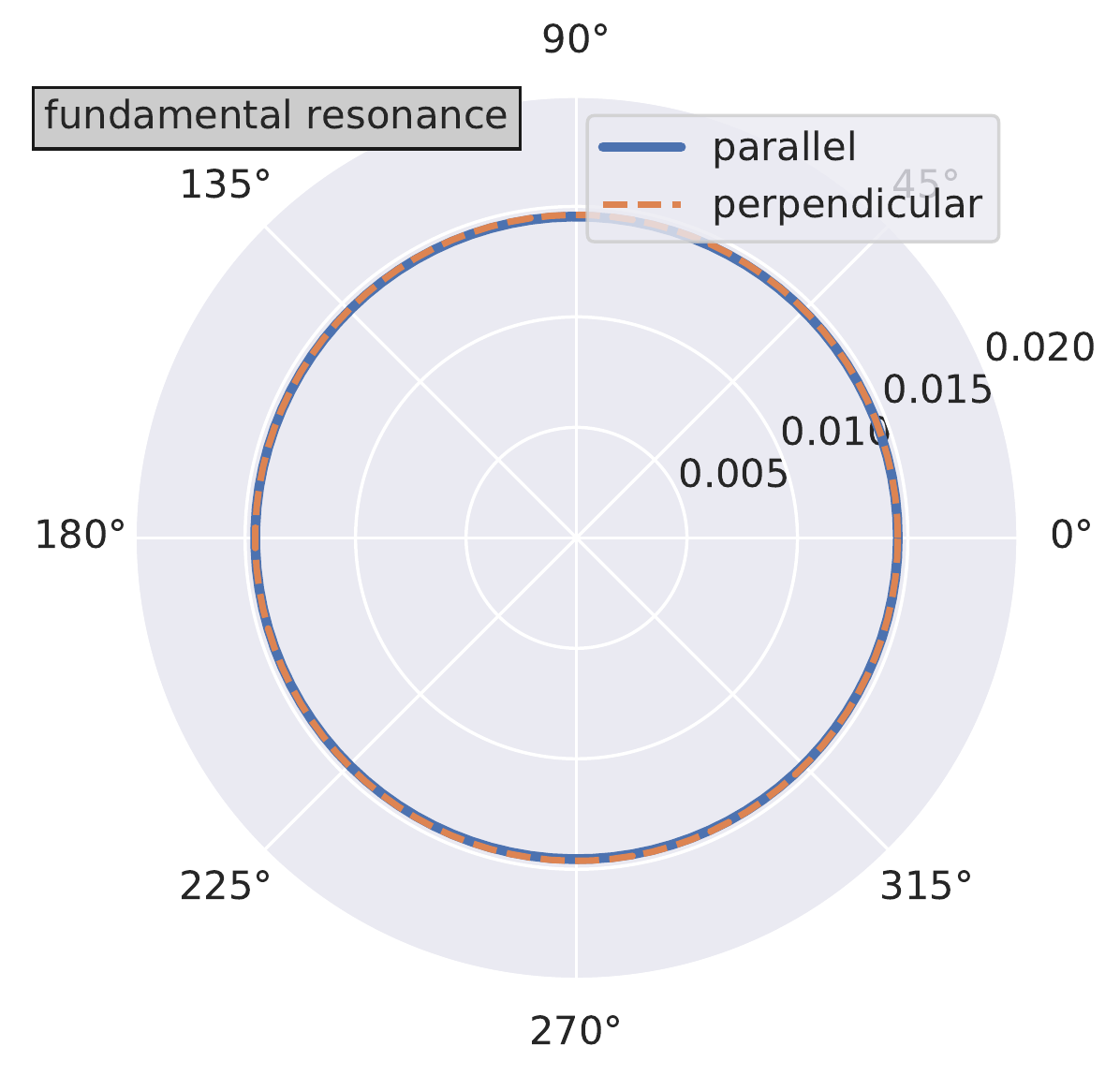}
	\includegraphics[width=.49\columnwidth]{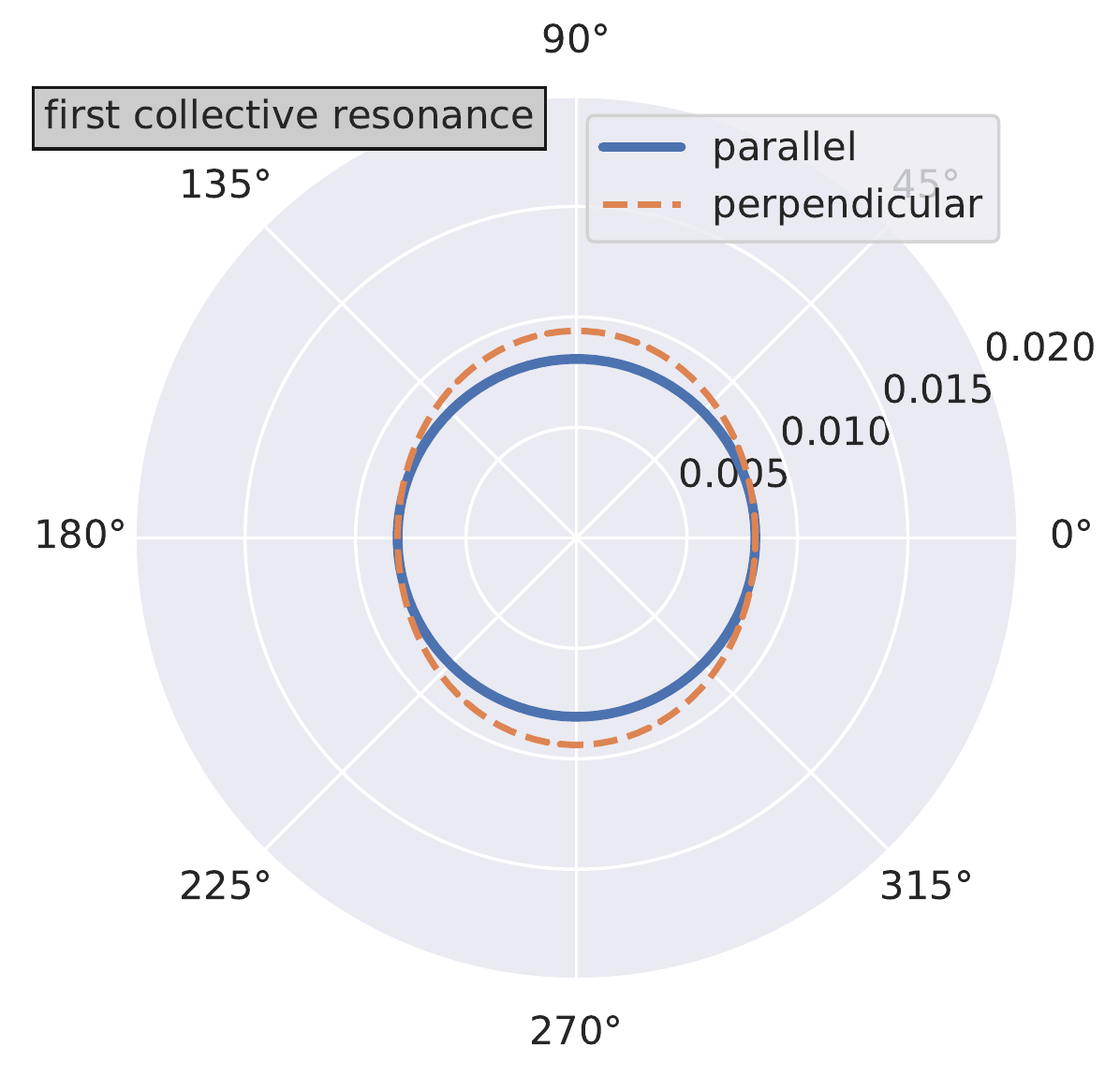}
	\caption{The scaled amplitude of the acoustic pressure, i.e., $R|p_\text{tot}|$, on the slices $x=0$ (parallel) and $z=0$ (perpendicular) at a distance $R=400a$ from the global origin. The array has $10 \times 10$ bubbles with a separation distance of eight radii. The normalized wavenumbers $ka=0.006755$ (left) and $ka=0.011261$ (right) correspond to the fundamental resonance and the first high-order collective mode, respectively.}
	\label{fig:field:polar}
\end{figure}

\paragraph{Far field at resonance}
The fields from the fundamental and collective resonances are clearly different close to the bubbles, but these differences rapidly diminish further from the array. Figure~\ref{fig:field:polar} shows the scaled magnitude of the field at a distance of 400~bubble radii. The fundamental resonance has a monopole structure with a spherically-symmetric field. The collective mode displays a slightly different form, with an elliptical shape in the plane perpendicular to the array, generated by the local hotspots. Furthermore, the far field from the collective mode is smaller, showing that modes other than the fundamental resonance do not tend to radiate to great distances.

\section{Conclusion}

The literature on physical experiments and mathematical models about the Minnaert resonance shows a downshift in the fundamental mode for small bubble ensembles, and an upshift for infinite arrays of bubbles. This work investigates the transition between these situations using a mass-spring model and the full-wave BEM solver. The modal analysis of the mass-spring model indeed predicts an initial downward shift in resonance frequency for small bubble ensembles, which reverses and becomes an upward frequency motion when more bubbles are present. More precisely, the bubble array must cover at least half a wavelength to reverse the downward frequency shifts. At the same time, higher-order collective resonances become significant, being excited more strongly by a perpendicular plane wave, and having a lower resonance frequency than the fundamental mode of bubbles oscillating in phase.

The accelerated full-wave BEM solver provides accurate frequency-response curves of square arrays with hundreds of bubbles. The linear computational complexity of the BEM facilitates the analysis of closely-packed bubble arrays that cover half a wavelength. The acoustic field at the surface of the bubbles confirms the identical phases of the bubble oscillations at the fundamental mode, and bubbles oscillating in opposite phases at the collective modes. These opposite phases generate local hotspots in acoustic pressure inside the array, which are not present in the monopole-type fundamental mode. Furthermore, the collective resonances do not radiate as far away as the spherically-symmetric field from the fundamental mode.

In conclusion, the numerical simulations using the accurate BEM solver of the Helmholtz equation confirm that bubble arrays covering more than half a wavelength have high-order collective modes dominating the fundamental resonance, thereby reversing the downward frequency shifts. Future research targets include improving BEM's computational efficiency, so that structures of several wavelengths can be modeled, and the resonance behavior of massive bubble arrays can be investigated.

\end{document}